\documentclass[a4paper,twocolumn,11pt,accepted=2026-03-24]{quantumarticle}
\pdfoutput=1
\usepackage[utf8]{inputenc}
\usepackage[T1]{fontenc}
\usepackage{tikz}
\usetikzlibrary{calc}
\usepackage{epsfig,amssymb,amsmath,physics,enumitem,qcircuit,bm}

\usepackage{array,multirow}
\usepackage{arydshln}
\usepackage{color,subfig}

\graphicspath{{./fig/}}

\definecolor{myurlcolor}{rgb}{0,0,0.7}
\definecolor{myrefcolor}{rgb}{0.8,0,0}
\usepackage{aliascnt}
\usepackage[unicode=true,pdfusetitle, bookmarks=true,bookmarksnumbered=false,bookmarksopen=false, breaklinks=false,pdfborder={0 0 0},backref=false,colorlinks=true, linkcolor=myrefcolor,citecolor=myurlcolor,urlcolor=myurlcolor]{hyperref}

\newtheorem{theorem}{Theorem}

\newaliascnt{lemma}{theorem}
\newtheorem{lemma}[lemma]{Lemma}
\aliascntresetthe{lemma}

\newaliascnt{corollary}{theorem}
\newtheorem{corollary}[corollary]{Corollary}
\aliascntresetthe{corollary}

\newtheorem{claim}[theorem]{Claim}

\def\equationautorefname~#1\null{%
  Eq.~(#1)\null
}

\newcommand{\eqsize}{\small}
\let\oldequation\equation
\let\oldendequation\endequation
\renewenvironment{equation}{\eqsize\oldequation}{\oldendequation}

\usepackage{accents}

\newcommand{\tpsi}{\tilde{\psi}}
\newcommand{\tPsi}{\tilde{\Psi}}

\newcommand{\qed}{\hfill$\square$}
\newcommand{\pf}[1]{\textit{Proof of \autoref{#1}.}}
\newcommand{\hf}{\frac{1}{2}}

\newcommand{\q}{\hat{q}}
\newcommand{\w}{\hat{w}}
\renewcommand{\v}{\hat{v}}

\newcommand{\inv}{{}^{-1}}

\newcommand{\bb}[1]{\left[ #1 \right]}
\newcommand{\cc}[1]{\left\{ #1 \right\}}

\newcommand{\appref}[1]{\hyperref[#1]{App.~\ref{#1}}}

\begin{document}

\title{Quantum Signal Processing and Quantum Singular Value Transformation on $U(N)$}

\author{Xi Lu}
\affiliation{School of Mathematical Science, Zhejiang University, Hangzhou, 310027, China}
\orcid{0000-0003-4121-3419}

\author{Yuan Liu}
\affiliation{Department of Electrical and Computer Engineering, North Carolina State University, Raleigh, NC 27606, USA}
\affiliation{Department of Computer Science, North Carolina State University, Raleigh, NC 27606, USA}
\orcid{0000-0003-1468-942X}

\author{Hongwei Lin}
\affiliation{School of Mathematical Science, Zhejiang University, Hangzhou, 310027, China}
\orcid{0000-0002-9337-9624}

\maketitle


\begin{abstract}
  Quantum signal processing and quantum singular value transformation are powerful tools to implement polynomial transformations of block-encoded matrices on quantum computers, and has achieved asymptotically optimal complexity in many prominent quantum algorithms.
  We propose a framework of quantum signal processing and quantum singular value transformation on $U(N)$, which realizes multiple polynomials simultaneously from a block-encoded input, as a generalization of those on $U(2)$ in the original frameworks.
  We provide a comprehensive characterization of achievable polynomial matrices and give recursive algorithms to construct the quantum circuits that realize desired polynomial transformations.
  As three example applications, we propose a framework to realize bi-variate polynomial functions, demonstrate $N$-interval decision achieving $\mathcal{O}(d)$ query complexity with a $\log_2 N$ improvement over iterative $U(2)$-QSP requiring $\mathcal{O}(d\log_2 N)$ queries, and present a quantum amplitude estimation algorithm achieving the Heisenberg limit without adaptive measurements.
\end{abstract}
\tableofcontents

\section{Introduction}

Quantum Signal Processing~(QSP) is a powerful tool for building quantum algorithms, capable of unifying many other existing algorithms~\cite{low2017optimal,gilyen2019quantum,martyn2021grand}.
QSP can be conceptualized as a framework of polynomial transformation of matrices, which maps a set of phase angles to a polynomial function to approximate a wide range of target functions.
Quantum Singular Value Transformation~(QSVT)~\cite{gilyen2019quantum}, another framework derived from QSP, extends the application to polynomial transformations of singular value of even non-square matrices.
QSP-based quantum algorithms have been developed for various tasks, such as Hamiltonian simulation~\cite{low2017optimal,low2019hamiltonian,ding2023simulating}, linear system solving~\cite{childs2017quantum}, ground state preparation~\cite{dong2022ground}, fixed-point quantum search~\cite{yoder2014fixed}.
QSP is also used to improve and simplify algorithms for quantum amplitude estimation~(QAE)~\cite{rall2023amplitude}, which is a fundamental task in quantum metrology~\cite{giovannetti2004quantum,giovannetti2006quantum,giovannetti2011advances} and has direct applications in numerical integration~\cite{montanaro2015quantum}, quantum tomography~\cite{haah2016sample,o2016efficient,aaronson2018shadow,hu2022logical,van2023quantum}, overlap and expectation value estimation in quantum simulation~\cite{knill2007optimal,kassal2008polynomial,kohda2022quantum,huggins2022nearly,simon2024amplified}, Gibbs sampling~\cite{van2019quantum}, variational quantum algorithms and quantum machine learning~\cite{peruzzo2014variational,wiebe2014quantum,wiebe2015quantum,kerenidis2019q}.
Recent research in QSP theories has focused on efficient realization of block encoding~\cite{li2023efficient,camps2024explicit}, classical evaluation of phase angles~\cite{chao2020finding,ying2022stable,dong2021efficient}, and generalization~\cite{laneve2023quantum,dong2024multi,low2024quantum,rossi2022multivariable,nemeth2023variants}.
Experiments have also been conducted to realize QSP on a noisy quantum computer~\cite{kikuchi2023realization,PhysRevApplied.23.034073}.

\begin{figure*}
    \centering
    \includegraphics[width=\linewidth]{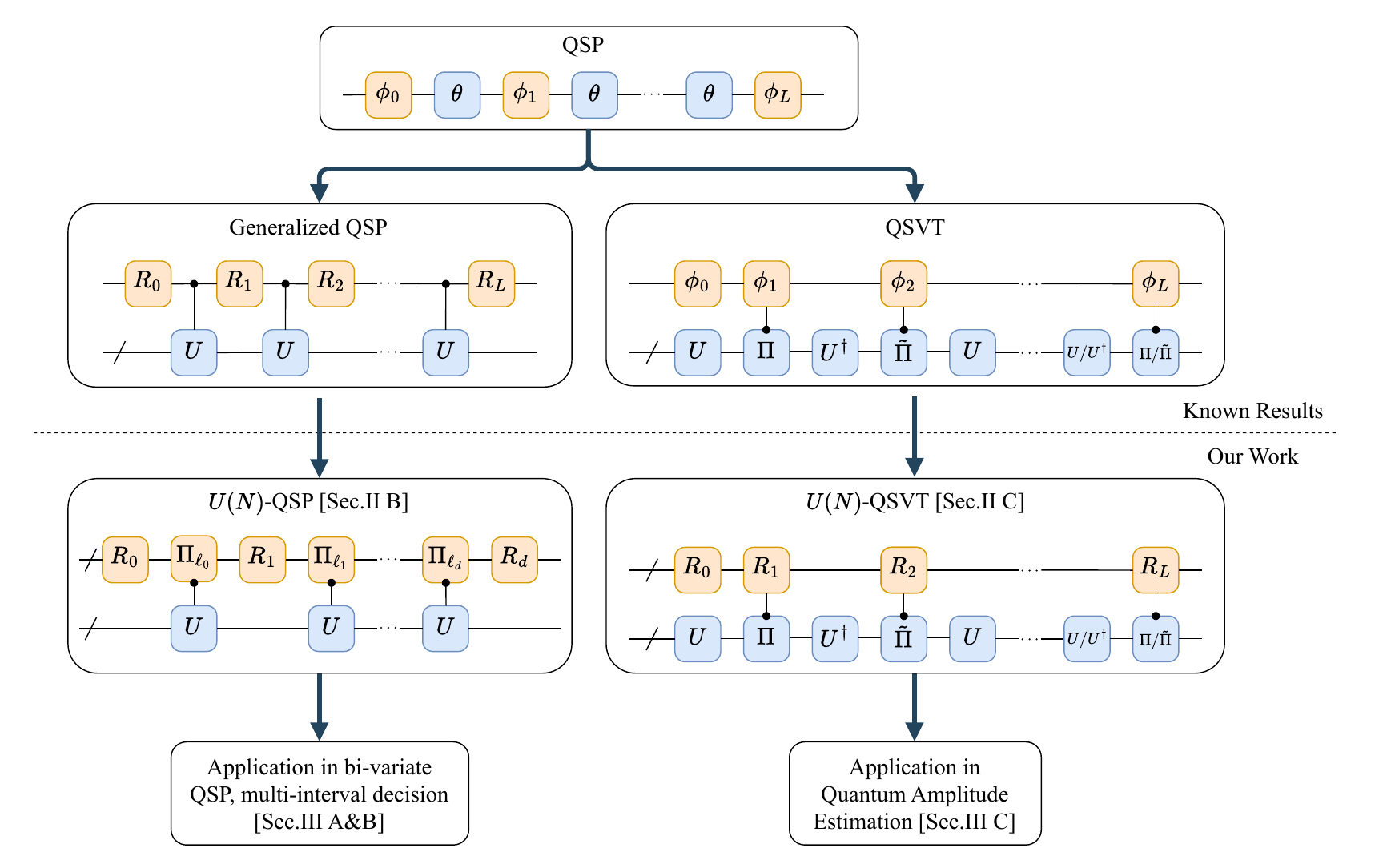}
    \caption{A summary of our contributions in the paper. The orange quantum gates are for parameterized unitaries or projectors, while the blue gates are for fixed input variables.}
    \label{fig:summary}
\end{figure*}

Meanwhile, the original framework of QSP has some restrictions that limit its applicability.
On the mathematical side, the original framework utilizes a series of parameterized $U(2)$ elements to realize a class of polynomial transformations, i.e., to construct a unitary transformation that is a block encoding of the target polynomial $P(U)$ given input $U$.
It is a natural question whether we can realize multiple polynomials at once if we use a sequence of parameterized $U(N)$ elements instead of $U(2)$ elements.
On the practical side, realizing multiple target functions simultaneously is essential for quantum algorithms like the quantum phase estimation~(QPE) and quantum amplitude estimation~(QAE) algorithms~\cite{brassard2000quantum}, where different measurement outcomes must correspond to different functions that collectively partition the domain of interest.
In addition, by expanding the toolkit in manipulating matrices in quantum computers, QSP and QSVT on $U(N)$ can also help us in more complicated tasks like the multi-variate generalization of QSP, which is much less understood than the uni-variate one, and known to have significant difficulties brought by its exponentially large target space and the commuting relations between different variables~\cite{rossi2022multivariable,nemeth2023variants}.

During the preparation of this paper, Laneve~\cite{laneve2023quantum} studied the generalization of QSP over $SU(N)$, which realizes state preparation for $N$-dimensional outputs by preparing the state $\sum_m P_m(z) \ket{m}$ from $\ket{0}$, with applications to quantum phase estimation.
Our framework on $U(N)$ is more general: Laneve's result corresponds to the special case of our \autoref{thm:unqsp_bk} where $\bm{P}(z)$ is a $N \times 1$ polynomial block, whereas we provide a complete theory for arbitrary $N \times N$ polynomial matrix blocks, enabling full exploitation of the $U(N)$ structure in all three applications demonstrated in this work.

In this paper, we establish a complete theory of QSP and QSVT on $U(N)$ with multiple outputs block encoded in a single unitary.
Given any mathematically permissible set of target polynomials, we provide a recursive procedure to determine the sequence of $U(N)$ circuit elements that realizes them.
Compared to the $U(2)$ framework, our $U(N)$ theory requires new perspectives on quantum circuits and additional mathematical tools from matrix theory.
We demonstrate three key applications: (i) bi-variate polynomial functions with an expanded range of achievable polynomials, (ii) multi-interval decision with an $O(\log_2 N)$ reduction in query complexity, and (iii) quantum amplitude estimation achieving the Heisenberg limit without adaptive measurements.
A graphical summary is given in \autoref{fig:summary}.

The paper is organized as follows.
\autoref{sec:theory} reviews fundamental results on uni-variate QSP with single block encoding, then generalizes to $U(N)$ for both QSP-U and QSVT, establishing the main theorems on achievable polynomial sets.
We then demonstrate three key applications where $U(N)$-QSP provides concrete advantages.
\autoref{sec:mqsp} presents bi-variate quantum signal processing that leverages product structures to realize bi-variate polynomial functions with practical advantages over naive decomposition approaches.
\autoref{sec:decision} demonstrates multi-interval decision with $\mathcal{O}(d)$ query complexity, providing a factor of $\log_2 N$ improvement over the $\mathcal{O}(d\log_2 N)$ complexity required by iterative $U(2)$-QSP methods.
\autoref{sec:qae} shows that QAE measurement outputs correspond to polynomial transformations on $U(N)$, achieving Heisenberg-limited precision in a single non-adaptive measurement round.
\autoref{sec:conclusion} provides conclusions and future research directions.

\section{Theories}\label{sec:theory}

In this section, we first briefly review the fundamental results about QSP in \autoref{sec:review_qsp}.
Then, in \autoref{sec:qsp_u} and \autoref{sec:qsvt}, we first define the generalization on $U(N)$, then construct a quantum circuit with parameterized parts that help achieving different target functions, and finally state and prove the achievable polynomial sets by the circuit.

\subsection{Review of Quantum Signal Processing and Quantum Singular Value Transformation}\label{sec:review_qsp}

To block-encode any matrix $A$ in a quantum operation, an ancilla system is used to construct a unitary $U$ such that,
\begin{equation}
    U\ket{\bm{0}}\ket{\psi} = \ket{\tilde{\bm{0}}} A\ket{\psi} + \cdots,
    \text{ or }
    U = \begin{bmatrix}
        A & * \\
        * & *
    \end{bmatrix},
	\label{eq:def-be}
\end{equation}
in which both $\ket{\bm{0}}$ and $\ket{\tilde{\bm{0}}}$ are qubits all set to zero, and we use different notations here to indicate that the number of qubits in them can be different, so that the block encoding can also be well defined for non-square matrix $A$.

In this paper, we focus on two algorithms in the QSP family, namely the QSP for unitary matrices and quantum singular value transformation~(QSVT) for general matrices.
In QSP-U, one use several controlled-$U$ operations to construct a block encoding of polynomials of $U$ of the form $P(U)=\sum_j c_j U^j$~\cite{dong2021efficient,motlagh2024generalized}.
A fundamental result in QSP-U is as follows.

\begin{theorem}[Generalized QSP~\cite{motlagh2024generalized}]\label{thm:qsp_u}
    Given any polynomial $P(z)$ of degree $d$ s.t. $|P(z)|\le 1, \forall |z|=1$. Then one can block-encode $P(U)$ using $d$ calls to controlled-$U$ for any unitary matrix input $U$.
\end{theorem}

In QSVT, however, one uses $U$ and $U^\dagger$ alternatively to construct a block encoding of singular-value polynomial transformations of $A$, which is defined as,
\begin{equation}
    P^{(SV)}(A) = \begin{cases}
		\sum_j P(\lambda_j) \dyad{\psi_j}{ \psi_j}, & \text{if } d \text{ is even}, \\
		\sum_j P(\lambda_j) \dyad{\psi_j}{\tpsi_j}, & \text{if } d \text{ is odd},
	\end{cases}
    \label{eq:def-svt}
\end{equation}
where $d$ is the number of calls to $U$ and $U^\dagger$ in total, and $A=\sum_j \lambda_j \dyad{\psi_j}{\tpsi_j}$ for two orthogonal sets $\{\ket{\psi_j}\},\{\ket{\tpsi_j}\}$ and $\lambda_j\in\mathbb{R}$, and $P$ naturally subjects to the parity condition that $P(-x)=(-1)^{d} P(x)$.
When $A$ is Hermitian, one can write $A=\sum_j \lambda_j \dyad{\psi_j}$ with $\lambda_j\in\mathbb{R}$, then the singular value polynomial transformation is equal to the matrix polynomial.
But in general they can be different.
A milestone result in the original framework of QSVT is as follows.

\begin{theorem}[QSVT~\cite{gilyen2019quantum}]\label{thm:qsvt}
    Given any polynomial $P(z)$ satisfying,
    \begin{enumerate}
        \item $\deg(P)\le d$;
        \item $P$ has parity $d \bmod 2$;
        \item $\forall x\in[-1,1]$, $|P(x)|\le 1$;
    \end{enumerate}
    and a general matrix $A$ block-encoded by a unitary $U$, one can block-encode $P^{(SV)}(A)$ using $d$ calls to $U$ and $U\inv$ in total.
\end{theorem}

Compared to QSP-U, it has inherent restrictions on parity, since singular value transformation~(SVT) by polynomials without definite parity is not well-defined in \autoref{eq:def-svt} and can give unexpected results.
One exception is that for Hermitian input, the SVT by polynomials without definite parity is well-defined since the left and right singular vector spaces share the same basis and is identical to the common polynomial transformation.
In this case SVT with complex-valued polynomials is also well-defined.
To tackle these two problems, we can utilize the linear combination of unitaries~(LCU)~\cite{childs2012hamiltonian}, given additional access to controlled $U$ and $U\inv$.

\subsection{$U(N)$-Quantum Signal Processing}\label{sec:qsp_u}

Given any unitary $U$ and complex polynomial matrix $\bm{P}(z)=\{P_{jk}(z)\}$, by $U(N)$-QSP we aim to construct the unitary transformation,
\begin{equation}
    \begin{bmatrix}
        P_{00}(U)    & P_{01}(U)    & \cdots & *      \\
        P_{10}(U)    & P_{11}(U)    & \cdots & *      \\
        \vdots       & \vdots       & \ddots & \vdots \\
        *            & *            & \cdots & *
    \end{bmatrix}.
    \label{eq:def_un_qsp}
\end{equation}

\begin{figure*}
    \centering
    \includegraphics[width=.7\linewidth]{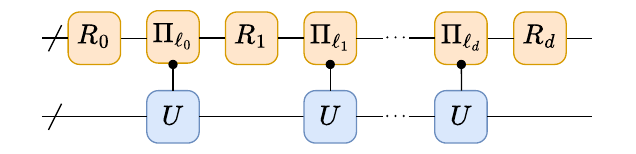}
    \caption{The quantum circuit for $U(N)$-QSP, parameterized by arbitrary unitaries $R_0, R_1, \ldots, R_d \in U(N)$ acting on an $N$-dimensional ancilla register. The circuit uses $d$ applications of controlled-$U$ operations, where $C_{\Pi_{\ell}}(U) := \sum_{j=0}^{\ell-1} \dyad{j}{j} \otimes U + \sum_{j=\ell}^{N-1} \dyad{j}{j} \otimes I$ applies $U$ when the ancilla is in computational basis states $\ket{0}$ through $\ket{\ell-1}$.}
    \label{fig:un_qsp}
\end{figure*}

For this task, we construct a quantum circuit in \autoref{fig:un_qsp}, parameterized by arbitrary unitary operators $R_0, R_1, \ldots, R_d \in U(N)$ acting on the ancilla register.
The circuit alternates between these unitaries and controlled-$U$ operations, where the control is based on computational basis projectors.
This structure generalizes the standard $U(2)$-QSP, which uses single-qubit unitaries
\begin{equation}
    R(\theta,\phi,\lambda) = \begin{bmatrix}
        e^{i(\lambda+\phi)} \cos\theta & e^{i\phi} \sin\theta \\
        e^{i\lambda} \sin\theta        & -\cos\theta
    \end{bmatrix},
\end{equation}
with $\dyad{1}$-controlled $U$ gates interleaved between them.

In the $U(N)$ generalization, we use $N$-dimensional ancilla systems where $N = 2^n$ for $n$ ancilla qubits.
The controlled operations are defined with respect to computational basis projectors: specifically, the $\Pi_{\ell}$-controlled $U$ gate is 
\begin{equation}
\begin{aligned}
C_{\Pi_{\ell}}(U) := \Pi_{\ell} \otimes U + (I - \Pi_{\ell}) \otimes I,
\end{aligned}
\end{equation}
where
\begin{equation}
\begin{aligned}
	\Pi_{\ell} := \sum_{j=0}^{\ell-1} \dyad{j}{j}.
\end{aligned}
\end{equation}
For example, $C_{\Pi_{N}}(U)$ is simply the uncontrolled $U$ gate, and $C_{\Pi_{N/2}}(U)$ is a single qubit controlled $U$ gate.

We characterize the achievable polynomials of $U(N)$-QSP in \autoref{fig:un_qsp} by the following lemmas and theorems.

\begin{lemma}[$U(N)$-QSP, forward]\label{thm:unqsp_fw}
	Using $d$ calls to a unitary $U$, the quantum circuit in \autoref{fig:un_qsp} implements the unitary operation,
	\begin{equation}
		\begin{bmatrix}
			P_{00}(U)    & P_{01}(U)    & \cdots & P_{0,N-1}(U) \\
			P_{10}(U)    & P_{11}(U)    & \cdots & P_{1,N-1}(U) \\
			\vdots       & \vdots       & \ddots & \vdots \\
			P_{N-1,0}(U) & P_{N-1,1}(U) & \cdots & P_{N-1,N-1}(U)
		\end{bmatrix}.
		\label{eq:mqsp_target}
	\end{equation}
	for a matrix of complex-valued polynomials $\{P_{jk}(z)\}$ of degrees no more than $d$, denoted as $\bm{P}(z)$.
\end{lemma}

\pf{thm:unqsp_fw}
The proof proceeds by induction on $d$.
For $d=0$, the circuit consists only of $R_0\otimes I$, so $P_{jk}$ is the constant function equal to the $(j,k)$-th entry of $R_0$.

For the inductive step, assume the lemma holds for $d-1$, meaning the circuit up to the $(d-1)$-th controlled-$U$ implements
$
	\bm{P}(U) = \sum_{l=0}^{d-1} \tilde{P}_l\otimes U^l
$
for some constant matrices $\{\tilde{P}_l\}$.
Applying $R_d$ on the ancilla register followed by $C_{\Pi_d}(U)$ gives
\begin{equation}
\begin{aligned}
	&
	(R_d \otimes I) C_{\Pi_d}(U) \bm{P}(U)
	\\ = &
	R_d \Pi_{\ell_d} \otimes U\bm{P}(U) + R_d (I - \Pi_{\ell_d}) \otimes \bm{P}(U),
\end{aligned}
\end{equation}
which expands to polynomials of degree no more than $d$ in $U$.
\qed

\begin{theorem}[$U(N)$-QSP, backward]\label{thm:unqsp_bk}
	Given any unitary $U$ and complex polynomial matrix $\bm{P}(z)$ of degrees no more than $d$, such that $\bm{P}(z)$ has all singular values in $[0,1]$ whenever $|z|\leq 1$.
	Then one can construct a quantum circuit with $d$ calls to controlled-$U$ to implement a block encoding of $\bm{P}(U)$, as defined in \autoref{eq:def_un_qsp}.
\end{theorem}

Before the proof of \autoref{thm:unqsp_bk}, we first prove a special case where $\bm{P}(z)$ is not a partial block of a bigger unitary, but the entire unitary matrix.

\begin{lemma}[$U(N)$-QSP for full unitary]\label{lem:unqsp_entire}
	Given any unitary $U$ and complex polynomial matrix $\bm{P}(z)$ that is unitary for all $|z|\leq 1$.
	Then one can determine the parameters $R_0, R_1, \ldots, R_d \in U(N)$ in \autoref{fig:un_qsp} to implement the unitary transformation $\bm{P}(U)$.
\end{lemma}

\pf{lem:unqsp_entire}
We use induction on $d$ to prove the lemma.

For $d=0$, each entry of \autoref{eq:mqsp_target} is constant, so we simply set $R_0$ to be the target unitary.
For the inductive step, assume the lemma holds for degree $d-1$. We show that for degree $d$, we can always find $R_d$ and a computational basis projector $\Pi_d = \sum_{j=0}^{\ell-1} \dyad{j}{j}$ such that $C_{\Pi_d}(U\inv) (R_d \otimes I) \bm{P}(U)$ is of the form \autoref{eq:mqsp_target} with degree no more than $d-1$.

Write
$
	\bm{P}(U) = \sum_{l=0}^{d} \tilde{P}_l\otimes U^l.
$
From the unitarity condition $\bm{P}(U)^\dagger \bm{P}(U) = I$, examining the coefficient of $U^d$ gives
$
	\tilde{P}_0^\dagger \tilde{P}_d = 0,
$
showing that the column spaces of $\tilde{P}_0$ and $\tilde{P}_d$ are orthogonal.
Let $r_{\tilde{P}_d}$ and $r_{\tilde{P}_0}$ be the ranks of $\tilde{P}_d$ and $\tilde{P}_0$ respectively. Then we have
\begin{equation}
\begin{aligned}
	r_{\tilde{P}_d} + r_{\tilde{P}_0} & \le N.
\end{aligned}
\end{equation}

Therefore, we can pick a basis $\{\ket{\psi_j}\}$ of $\mathbb{C}^N$ such that the first $r_{\tilde{P}_d}$ basis vectors span the column space of $\tilde{P}_d$, and the last $r_{\tilde{P}_0}$ basis vectors span the column space of $\tilde{P}_0$. 
Picking any 
\begin{equation}\label{eq:pick_ell}
\begin{aligned}
r_{\tilde{P}_d} \le \ell_d \le N - r_{\tilde{P}_0},
\end{aligned}
\end{equation}
and let
\begin{equation}\label{eq:Rd}
\begin{aligned}
	&
	R_d = \sum_{j=0}^{N-1} \dyad{\psi_j}{j}.
\end{aligned}
\end{equation}

Then,
\begin{equation}
\begin{aligned}
	&
	C_{\Pi_d}(U\inv) (R_d\inv \otimes I) \bm{P}(U)
	\\ = &
	\sum_{l=0}^{d} \bb{
		\Pi_d R_d\inv \tilde{P}_l\otimes U^{l-1} + (I-\Pi_d) R_d\inv \tilde{P}_l\otimes  U^l
	}
	\\ = &
	\sum_{l=0}^{d-1} \left[
		\Pi_d R_d\inv \tilde{P}_{l+1} + (I-\Pi_d) R_d\inv \tilde{P}_{l}
	\right]\otimes U^{l},
\end{aligned}
\label{eq:recur}
\end{equation}
where in the last equation the degree-$d$ term vanishes since $(I - \Pi_d) R_d\inv \tilde{P}_d = 0$, and the degree-$(-1)$ term vanishes since $\Pi_d R_d\inv \tilde{P}_0 = 0$.

By induction, this provides a constructive algorithm to determine $R_0, R_1, \ldots, R_d$.
\qed

\pf{thm:unqsp_bk}
Since $I - \bm{P}(z)^\dagger\bm{P}(z)$ is positive semidefinite on $|z|=1$, by the \textit{Polynomial Matrix Spectral Factorization Theorem}~\cite{wiener1957prediction,ephremidze2014elementary}, there is a polynomial matrix $\bm{Q}(z)$ of degree no more than $d$ such that,
\begin{eqnarray}
	I - \bm{P}(z)^\dagger\bm{P}(z) &=& \bm{Q}(z)^\dagger\bm{Q}(z).
    \label{eq:ippqq}
\end{eqnarray}

Next, we construct a polynomial matrix block $\bm{R}(z)$ such that,
\begin{equation}
    \begin{bmatrix}
        \bm{P}(z) & \multirow{2}{*}{$\bm{R}(z)$} \\ 
        \bm{Q}(z) & \\
    \end{bmatrix}
\label{eq:pqr}
\end{equation}
is unitary on $|z|=1$, where $\bm{R}(z)$ has proper size to make it a square matrix and satisfies,
\begin{equation}
\begin{aligned}
    &
    \begin{bmatrix}
            \bm{P}(z) & \multirow{2}{*}{$\bm{R}(z)$} \\ 
            \bm{Q}(z) & \\
    \end{bmatrix}
    \begin{bmatrix}
            \bm{P}(z)^\dagger & \bm{Q}(z)^\dagger \\
            \multicolumn{2}{c}{\bm{R}(z)^\dagger}
    \end{bmatrix}
    \\ = &
    \begin{bmatrix}
		\bm{P}(z) \\ \bm{Q}(z)
	\end{bmatrix}
    \begin{bmatrix}
		\bm{P}(z)^\dagger & \bm{Q}(z)^\dagger
	\end{bmatrix}
    +
    \bm{R}(z) \bm{R}(z)^\dagger
    =
    I,
\end{aligned}
\end{equation}

Again, this is always possible since
\begin{equation}
	I - 
    \begin{bmatrix}
		\bm{P}(z) \\ \bm{Q}(z)
	\end{bmatrix}
    \begin{bmatrix}
		\bm{P}(z)^\dagger & \bm{Q}(z)^\dagger
	\end{bmatrix}
	\label{eq:i_pq}
\end{equation}
is positive semidefinite for all $|z|=1$, as
\begin{equation}
    \begin{bmatrix}
        \bm{P}(z)^\dagger & \bm{Q}(z)^\dagger
    \end{bmatrix}
    \begin{bmatrix}
        \bm{P}(z) \\ \bm{Q}(z)
    \end{bmatrix}
    = I
\end{equation}
from \autoref{eq:ippqq} implies that 
\begin{equation}
    \begin{bmatrix}
        \bm{P}(z) \\ \bm{Q}(z)
    \end{bmatrix}
    \begin{bmatrix}
        \bm{P}(z)^\dagger & \bm{Q}(z)^\dagger
    \end{bmatrix}
\end{equation}
is identity in some subspace.
Finally,
\begin{equation}
    \begin{bmatrix}
        \bm{P}(U) & \multirow{2}{*}{$\bm{R}(U)$} \\ 
        \bm{Q}(U) & \\
    \end{bmatrix}
\end{equation}
is a block encoding of $\bm{P}(U)$ and by \autoref{lem:unqsp_entire}, it can be implemented as desired.
\qed

\autoref{thm:unqsp_bk} is a generalization of the results in \cite{motlagh2024generalized}, in which only one ancilla qubit is used, and the corresponding $\bm{P}(z)$ contains a single entry $p(z)$, with prerequisites $|p(z)|\leq 1$ for all $|z|=1$.

To find the circuit parameters in \autoref{thm:unqsp_bk} given only the $\bm{P}(z)$ block, it is sufficient to find only $\bm{Q}(z)$ in \autoref{eq:ippqq} using the algorithm described by the constructive proof in \cite{ephremidze2014elementary} and ignore the $\bm{R}(z)$ block.
To see this, we merge $\bm{Q}(z)$ into $\bm{P}(z)$ and write
\begin{equation}
	\bm{P}(z) = \sum_{l=0}^{d} \begin{bmatrix}
		\tilde{P}_l & *
	\end{bmatrix} z^l,
\end{equation}
where $*$ corresponds to the $\bm{R}(z)$ block in \autoref{eq:pqr} that we no longer need here.

Similar to the proof of \autoref{lem:unqsp_entire}, the constraint $\bm{P}(U)^\dagger\bm{P}(U)=I$ gives
\begin{equation}
	\begin{bmatrix}
		\tilde{P}_0^\dagger \\ *
	\end{bmatrix}
	\begin{bmatrix}
		\tilde{P}_d & *
	\end{bmatrix} = 0,
\end{equation}
which implies that $\tilde{P}_0^\dagger \tilde{P}_d = 0$.
Let $M$ be the number of columns of $\bm{P}(z)$, then there is a $M\times M$ projector $\tilde{\Pi}_d$ such that $\tilde{\Pi}_d \tilde{P}_0 = (I-\tilde{\Pi}_d) \tilde{P}_d = 0$.
One can choose,
\begin{equation}
	\Pi_d = \begin{bmatrix}
		\tilde{\Pi}_d & 0 \\
		0 & 0
	\end{bmatrix},
\end{equation}

Then \autoref{eq:recur} becomes,
\begin{equation}
\begin{aligned}
	&
	\sum_{l=0}^{d-1} \biggl(
		\Pi_d\begin{bmatrix}
			\tilde{P}_{l+1} & *
		\end{bmatrix} + (I-\Pi_d)\begin{bmatrix}
			\tilde{P}_{l} & *
		\end{bmatrix}
	\biggr) \otimes U^{l}
	\\ = &
	\sum_{l=0}^{d-1} 
	\begin{bmatrix}
		\tilde{\Pi}_d\tilde{P}_{l+1} + (I-\tilde{\Pi}_d)\tilde{P}_{l} & *
	\end{bmatrix}
	\otimes U^l.
\end{aligned}
\end{equation}
So it is sufficient to determine $\Pi_d$ recursively only using information from the selected columns (or equivalently, rows) instead of the whole matrix.

When the target block encoded matrix $\bm{P}(z)$ has rank no more than $N/2$, we have $r_{\tilde{P}_0}, r_{\tilde{P}_d} \le N/2$ in \autoref{eq:pick_ell}, so we can always choose $\ell_d = N/2$, and simplify the circuit in \autoref{fig:un_qsp} to use only single-qubit controlled-$U$ gates.

In some cases, we may need to realize Laurent polynomials~\cite{sinanan2024single}, where each entry is of the form $P_{jk}(z) = \sum_{l=-d}^{d} P_{jk,l} z^l$.

\begin{corollary}[$U(N)$-QSP for Laurent polynomials]\label{cor:un_qsp_laurent}
	Given any unitary $U$ and Laurent polynomial matrix $\bm{P}(z)$ of degrees no more than $d$, such that $\bm{P}(z)$ has all singular values in $[0,1]$ whenever $|z|\leq 1$.
	Then one can construct a quantum circuit with $(2d)$ calls to the double-headed gate $C_{\Pi}(U^{1/2}, U^{-1/2}):=\Pi\otimes U^{1/2} + (I-\Pi)\otimes U^{-1/2}$ to implement a block encoding of $\bm{P}(U)$.
\end{corollary}

\pf{cor:un_qsp_laurent}
We construct the circuit to realize the degree-$(2d)$ polynomial matrix $\sum_{l=-d}^{d} P_{jk,l} z^{l+d}$ using \autoref{thm:unqsp_bk}.
By replacing each controlled-$U$ gate with the double-headed gate $C_{\Pi}(U^{1/2}, U^{-1/2})$, we realize the Laurent polynomial matrix $\bm{P}(z)$ as desired.
\qed

\begin{corollary}[Approximate $U(N)$-QSP]\label{coro:qsp_aprx}
	Given any unitary $U$ and complex polynomial matrix $\bm{P}(z)$ of degrees no more than $d$, such that $\bm{P}(z)$ has all singular values in $[0,1]$ whenever $|z|\leq 1$.
	Let $\mathcal{G}$ be a finite set of unitary gates with determinant one and their inverses, generating a dense subset of $SU(N)$, where $N$ is the ancilla dimension.
	Then for any $\epsilon > 0$, one can construct a quantum circuit using gates from $\mathcal{G}$ with $d$ calls to controlled-$U$ that implements a unitary $\tilde{\bm{U}}$ such that the realized block-encoded matrix satisfies
	\begin{equation}
		\left\|\Pi^{\dagger} \tilde{\bm{U}} \Pi - \bm{P}(U)\right\| \leq \epsilon,
	\end{equation}
	where $\Pi = |0\ldots 0\rangle\langle 0\ldots 0|$ projects onto the ancilla subspace.
	The total number of gates from $\mathcal{G}$ required is
	\begin{equation}
		O\left(d \cdot \log^c\left(\frac{d}{\epsilon}\right)\right),
	\end{equation}
	where $c\geq 1$ is a constant depending on the gate set $\mathcal{G}$, and the constant hidden in the $O(\cdot)$ notation depends polynomially on the ancilla dimension $N$.
\end{corollary}

\pf{coro:qsp_aprx}
By \autoref{thm:unqsp_bk}, we construct the circuit with unitaries $R_0, R_1, \ldots, R_d$ as in \autoref{fig:un_qsp}.

Let $\bm{U}_{\text{exact}} = R_0 \cdot C_{\Pi_1}(U) \cdot R_1 \cdots R_d \cdot C_{\Pi_d}(U)$ be the exact circuit by \autoref{thm:unqsp_bk}.
Each $R_k$ is a unitary operator on the ancilla system of dimension $N$.
The Solovay-Kitaev theorem ~\cite{dawson2005solovay,pham2013optimization,kuperberg2023breaking} of dimension $N$ ensures that for any $V\in SU(N)$, there exists a finite sequence $S$ of gates from $\mathcal{G}$ of length $O(\log^c(1/\delta))$ such that $\|V - S\| < \delta$, where $c$ is a constant depending on $\mathcal{G}$ and the constant in the $O(\cdot)$ notation depends polynomially on the ancilla dimension $N$.
We apply this to approximate each $R_k$ by $\tilde{R}_k$ with error tolerance $\delta_k$, obtaining
\begin{equation}
	\|R_k - \tilde{R}_k\| \leq \delta_k.
\end{equation}

We choose $\delta_k = \frac{\epsilon}{2d+2}$ for all $k \in \{0, 1, \ldots, d\}$.
The compiled circuit is then
\begin{equation}
	\tilde{\bm{U}} = \tilde{R}_0 \cdot C_{\Pi_1}(U) \cdot \tilde{R}_1 \cdots \tilde{R}_d \cdot C_{\Pi_d}(U).
\end{equation}

We have,
\begin{equation}\label{eq:unitary-diff}
\begin{aligned}
	&
	\norm{ \Pi^{\dagger} \tilde{\bm{U}} \Pi - \bm{P}(U) } \\
	=&
	\norm{ \tilde{\bm{U}} - \bm{U}_{\text{exact}} } \\
	=&
	\left\| \tilde{R}_0 \cdot C_{\Pi_1}(U) \cdot \tilde{R}_1 \cdots \tilde{R}_d \cdot C_{\Pi_d}(U) \right. \\
	& \left. - R_0 \cdot C_{\Pi_1}(U) \cdot R_1 \cdots R_d \cdot C_{\Pi_d}(U) \right\|
	\\ \le &
	\sum_{k=0}^{d} \| \tilde{R}_k - R_k \| \\
	\leq & \epsilon,
\end{aligned}
\end{equation}
where we used
\begin{equation}
	\|A_1 A_2 \cdots A_n - B_1 B_2 \cdots B_n\| \leq \sum_{j=1}^n \|A_j - B_j\|,
\end{equation}
for any unitary operators $\{A_j, B_j\}$.

Each approximation $\tilde{R}_k$ requires $O(\log^c(1/\delta_k))$ gates from $\mathcal{G}$.
Since there are $d+1$ unitaries $R_k$ to approximate (for $k = 0, 1, \ldots, d$) with $\delta_k = \frac{\epsilon}{2d+2}$, the total gate count is
\begin{equation}
\begin{aligned}
	(d+1) \cdot O\left(\log^c\left(\frac{2d+2}{\epsilon}\right)\right)
	&= O\left(d \cdot \log^c\left(\frac{d}{\epsilon}\right)\right),
\end{aligned}
\end{equation}
where the constant hidden in the $O(\cdot)$ notation depends polynomially on the ancilla dimension $N$.
\qed

\subsection{$U(N)$-Quantum Singular Value Transformation}\label{sec:qsvt}

\begin{figure*}
	\centering
    \includegraphics[width=.7\textwidth]{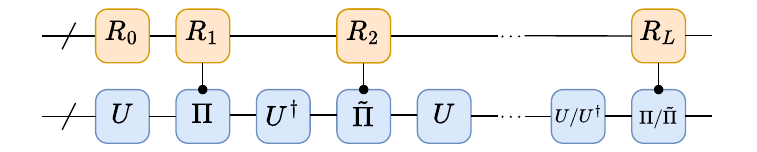}
    \caption{The $U(N)$-QSVT unit, in which $\Pi=\dyad{\bm{0}}$ and $\tilde{\Pi}=\dyad{\tilde{\bm{0}}}$. The $U$ and $U^\dagger$ gates applied to the second register alternate, and it depends on the parity of $d$ whether the last two gates in the second register are $U$ and $\Pi$, or $U^\dagger$ and $\tilde{\Pi}$.}
    \label{fig:mqsvt_unit}
\end{figure*}

In this subsection we assume all polynomial transformations of matrices are the singular value polynomial transformations in \autoref{eq:def-svt}, and without ambiguity we omit the superscript $(SV)$.
Assume $\ket{\psi}$ is exactly some right singular vector $\ket{\psi_k}$ of $A$.
Define $\ket{\Psi_k}=\ket{\bm{0}}\ket{\psi_k}$, $\ket{\tPsi_k}=\ket{\tilde{\bm{0}}}\ket{\tpsi_k}$, and define $\ket{\Psi_k^\perp},\ket{\tPsi_k^\perp}$ by
\begin{eqnarray}
	U\ket{\Psi_k} &=& \lambda_{k}\ket{\tPsi_k} + \bar{\lambda}_{k}\ket{\tPsi_k^\perp},\\
	U^\dagger\ket{\tPsi_k} &=& \lambda_{k}\ket{\Psi_k} - \bar{\lambda}_{k}\ket{\Psi_k^\perp},
\end{eqnarray}
where $\bar{\lambda}_{k}:=\sqrt{1-\lambda_{k}^2}$.
Thus in the basis $(\ket{\Psi_k},\ket{\Psi_k^\perp})\to(\ket{\tPsi_k},\ket{\tPsi_k^\perp})$,
\begin{equation}
	U = \begin{bmatrix}
		\lambda_{k} & -\bar{\lambda}_{k} \\
		\bar{\lambda}_{k} & \lambda_{k}
	\end{bmatrix}.
\end{equation}

Given a general matrix $A$ and a matrix of polynomials $\bm{P}$, the $U(N)$-QSVT is defined as the unitary transformation,
\begin{equation}
\begin{aligned}
    &
    \sum_j \ket{j} \ket{\bm{0}} \ket{\phi_j}
    \\ \mapsto &
    \sum_{k} \left[
        \ket{k} \ket{\bm{0}} \sum_j P_{kj}(A)\ket{\phi_j} + \ket{\bm{0}^\perp} \ket{\cdots}
    \right].
\end{aligned}
\label{eq:def_un_qsvt}
\end{equation}

Similar to the idea of qubitization~\cite{gilyen2019quantum}, we first give the following two lemmas that work with one singular value $\lambda_{k}$.

\begin{lemma}[$U(N)$-QSVT for one singular value]\label{lem:un_qsvt_lambda}
    If $d$ is odd, then the quantum circuit in \autoref{fig:mqsvt_unit} implements the unitary transformation,
	\begin{equation}
		\begin{aligned}
			&
			\ket{j} \ket{\Psi_k}
			\\ \mapsto &
			\sum_{k} \ket{k} \left[
                P_{kj}(\lambda_{k}) \ket{\tPsi_k} + \bar{\lambda}_{k} Q_{kj}(\lambda_{k}) \ket{\tPsi_k^\perp}
            \right],
		\end{aligned}
		\label{eq:mqsvt_unit_fw}
	\end{equation}
    for some $d$-polynomials $\{P_{kj}\}$ and $(d-1)$-polynomials $\{Q_{kj}\}$ such that,
    \begin{equation}
        \sum_k \left[
            |P_{kj}(x)|^2 + (1-x^2) |Q_{kj}(x)|^2
        \right] \equiv 1.
    \end{equation}
    Otherwise, if $d$ is even, then in \autoref{eq:mqsvt_unit_fw} the $\ket{\tPsi_k},\ket{\tPsi_k^\perp}$ should be replaced by $\ket{\Psi_k},\ket{\Psi_k^\perp}$.
\end{lemma}

\pf{lem:un_qsvt_lambda}
We prove by induction on $d$.
For $d=0$, the output state is simply $\sum_k u_{kj} \ket{k}\ket{\Psi_k}$, with $u_{kj}$ being the $(k,j)$-th entry of $R_0$, and these constant functions are 0-polynomials.

Suppose the lemma holds for some even number $(d-1)$, i.e., the state before the final $U$ and $C_{\Pi}(R_d)$ gates in \autoref{fig:mqsvt_unit} is \autoref{eq:mqsvt_unit_fw}.
Then after applying the two gates, the state is,
\begin{equation}
\begin{aligned}
	\sum_k \ket{k} & \biggl\{
		\sum_l u_{kl} 
		\left[
			\lambda_{k} P_{lj}(\lambda_{k}) - (1-\lambda_{k}^2) Q_{lj}(\lambda_{k})
		\right] \ket{\tPsi_k}
	\\ & +\bar{\lambda}_{k} \left[
			P_{lk}(\lambda_{k}) + \lambda_{k} Q_{kj}(\lambda_{k})
		\right] \ket{\tPsi_k^\perp}
	\biggr\},
\end{aligned}
\end{equation}
which is of the desired form in \autoref{eq:mqsvt_unit_fw} with polynomials satisfying both the degree and parity constraints.

The case when $d$ is even is analogous.
\qed

By the linearity of quantum circuits, the single singular value case can be immediately generalized as follows.

\begin{lemma}[$U(N)$-QSVT, forward]\label{thm:unqsvt_fw}
	The quantum circuit in \autoref{fig:mqsvt_unit} implements the unitary transformation \autoref{eq:def_un_qsvt} for some matrix of polynomials $\bm{P}$.
\end{lemma}

The main theorem showing the usefulness of the quantum circuit in \autoref{fig:mqsvt_unit}, as a generalization of \autoref{thm:qsvt}, is as follows.

\begin{theorem}[$U(N)$-QSVT, backward]\label{thm:unqsvt_bk}
    Given a matrix $A$ block-encoded by $U$ as in \autoref{eq:def-be}, and a polynomial matrix $\bm{P}(x)$ such that $I-\bm{P}(x)^\dagger \bm{P}(x)$ is positive semidefinite for all $x\in[-1,1]$, with $d$ calls to $U$ and $U\inv$ in total, one can implement a block encoding of $\bm{P}(A)$ by the following unitary transformation,
    \begin{equation}
        \begin{aligned}
            &
            \ket{0} \sum_{j} \ket{j} \ket{\bm{0}} \ket{\phi_j}
            \\ \mapsto &
            \ket{0} \sum_{k} \ket{k} \ket{\bm{0}} \sum_{j} P_{kj}(A) \ket{\phi_{j}} + \ket{1} \ket{\cdots}.
        \end{aligned}
        \label{eq:mqsvt}
    \end{equation}
\end{theorem}

\begin{lemma}[$U(N)$-QSVT from 2 polynomials]\label{lem:mqsvt_unit_bw}
    Given a matrix $A$ and its block encoding $U$, a matrix of $d$-polynomials $\bm{P}(x)$ and a matrix of $(d-1)$-polynomials $\bm{Q}(x)$ of the same size such that,
	\begin{equation}
		\bm{P}(x)^\dagger \bm{P}(x) + (1-x^2) \bm{Q}(x)^\dagger \bm{Q}(x) \equiv
		I,
		\label{eq:mqsvt_constr}
	\end{equation}
	for all $x\in[-1,1]$, one can find $R_0,\cdots,R_d$ in \autoref{fig:mqsvt_unit} to make it implement the transformation \autoref{eq:mqsvt_unit_fw} for each $j$.
\end{lemma}

\pf{lem:mqsvt_unit_bw}
We prove by induction on $d$.
The case $d=0$ is trivial, as $\bm{Q}(x)=0$ and $\bm{P}(x)$ is a constant unitary matrix, and one can simply let $R_0=\bm{P}(x)$.

Suppose the lemma holds for some even $(d-1)$, and now we consider the case for $d$.
Write,
\begin{eqnarray}
	\bm{P}(x) &=& \sum_{l=0}^{(d-1)/2} \tilde{P}_{2l+1} x^{2l+1}, \label{eq:p_series}\\
	\bm{Q}(x) &=& \sum_{l=0}^{(d-1)/2} \tilde{Q}_{2l} x^{2l}.
\end{eqnarray}

Picking the $x^{2d}$ terms out of the constraint \autoref{eq:mqsvt_constr},
\begin{equation}
	\tilde{P}_{d}^\dagger \tilde{P}_{d} - \tilde{Q}_{d-1}^\dagger \tilde{Q}_{d-1} = 0,
\end{equation}
so there is a unitary $R_d$ such that
$
R_d^\dagger \tilde{P}_{d} = \tilde{Q}_{d-1}
$.

Write $R_d^\dagger = \{u_{kl}\}$.
Then,
\begin{equation}
\begin{aligned}
	&
	(I\otimes U)\inv C_{\Pi}(R_d)\inv \cdot
	\\ &
	\sum_{k} \ket{k} \left[
		P_{kj}(\lambda_{k}) \ket{\tPsi_k} + \bar{\lambda}_{k} Q_{kj}(\lambda_{k}) \ket{\tPsi_k^\perp}
	\right]
	\\ = &
	\sum_k \ket{k} \left\{
		\left[
			\sum_l u_{kl} \lambda_{k} P_{lj}(\lambda_{k}) + (1-\lambda_{k}^2) Q_{kj}(\lambda_{k})
		\right] \ket{\Psi_k}
	\right.
	\\ & \left. +\bar{\lambda}_{k} \left[
			-\sum_l u_{kl} P_{lj}(\lambda_{k}) + \lambda_{k} Q_{kj}(\lambda_{k})
		\right] \ket{\Psi_k^\perp}
	\right\},
\end{aligned}
\end{equation}
in which the coefficient polynomial of $\ket{\Psi_k}$ is actually a $(d-1)$-polynomial, since its $\lambda_{k}^{d+1}$ term coefficient $\sum_l
u_{kl} (\tilde{P}_d)_{lj} - (\tilde{Q}_{d-1})_{kj} = 0$, and similarly the coefficient polynomial of $\ket{\Psi_k^\perp}$ is actually a $(d-2)$-polynomial.
So we reduce the degree of the problem by 1.

The case when $d$ is even is analogous.
\qed

\pf{thm:unqsvt_bk}
All we need to show is that one can find a matrix of $d$-polynomials $\bm{P}_1(x)$ and a matrix of $(d-1)$-polynomials $\bm{Q}_1(x)$ such that,
\begin{equation}
	\begin{bmatrix}
		\bm{P}(x)^\dagger & \bm{P}_1(x)^\dagger
	\end{bmatrix}
	\begin{bmatrix}
		\bm{P}(x) \\ \bm{P}_1(x)
	\end{bmatrix}
	+
	(1-x^2) \bm{Q}_1(x)^\dagger \bm{Q}_1(x)
	=
	I,
	\label{eq:mqsvt_expansion_target}
\end{equation}
such that by rearranging order, one can label the flag qubit corresponding to the $\bm{P}(x)$ block to zero while $\bm{P}_1(x)$ and $\bm{Q}_1(x)$ to one, to obtain the desired block encoding of $\bm{P}(A)$.

Again, we prove the case when $d$ is even, and the other case is analogous.
Write $\bm{P}$ as \autoref{eq:p_series}.
Make substitution $x\to\cos\frac{\theta}{2}$, then
$
	\bm{P}(x)
	=
	e^{-i\frac{d\theta}{2}}
	\tilde{\bm{P}}(e^{i\theta}),
$
for some polynomial matrix $\tilde{\bm{P}}(z)$ of degree no more than $d$.
Moreover,
$
	I - \tilde{\bm{P}}(e^{i\theta})^\dagger \tilde{\bm{P}}(e^{i\theta})
$
is positive semidefinite for all $|z|=1$.
By the \textit{Polynomial Matrix Spectral Factorization Theorem}~\cite{wiener1957prediction,ephremidze2014elementary}, there is a polynomial matrix $\tilde{\bm{Q}}(e^{i\theta})$ of degree no more than $d$ such that
\begin{equation}
	I - \tilde{\bm{P}}(e^{i\theta})^\dagger \tilde{\bm{P}}(e^{i\theta})
	=
	\tilde{\bm{Q}}(e^{i\theta})^\dagger \tilde{\bm{Q}}(e^{i\theta}).
\end{equation}
Write
\begin{equation}
	e^{-i\frac{d\theta}{2}}
	\tilde{\bm{Q}}(e^{i\theta})
	=
	\bm{P}_1\left(\cos\frac{\theta}{2}\right)
	+
	\sin\frac{\theta}{2}
	\bm{Q}_1\left(\cos\frac{\theta}{2}\right),
\end{equation}
then $\bm{P}_1(x)$ is a matrix of $d$-polynomials and $\bm{Q}_1(x)$ is a matrix of $(d-1)$-polynomials.
Since,
\begin{equation}
\begin{aligned}
	&
	\bm{Q}(e^{i\theta})^\dagger
	\bm{Q}(e^{i\theta})
	-
	\bm{P}_1(x)^\dagger
	\bm{P}_1(x)
	-
	(1-x^2)
	\bm{Q}_1(x)^\dagger
	\bm{Q}_1(x)
	\\ = &
	\sin\frac{\theta}{2} \left[
		\bm{P}_1(x)^\dagger
		\bm{Q}_1(x)
		+
		\bm{Q}_1(x)^\dagger
		\bm{P}_1(x)
	\right],
\end{aligned}
\end{equation}
in which the left hand side is even about $\theta$ and the right hand side is odd, thus both are zero.
As a result, \autoref{eq:mqsvt_expansion_target} holds.
Finally, the proof is completed by \autoref{lem:mqsvt_unit_bw}.
\qed

Like the original QSVT algorithm, for Hermitian matrix input $A$, one can block-encode polynomial matrix $\bm{P}(A)$ without definite parity constraints, by splitting the polynomial into even and odd parts, namely $\bm{P}_e(A)$ and $\bm{P}_o(A)$ such that $\bm{P}(A)=\frac{1}{2}(\bm{P}_e(A)+\bm{P}_o(A))$, and using \textit{Linear Combination of Unitaries} (LCU)~\cite{childs2017quantum} to obtain a block encoding of $\bm{P}(A)$.
To guarantee the nonnegativity of $I-\bm{P}_e(A)^\dagger\bm{P}_e(A)$ and $I-\bm{P}_o(A)^\dagger\bm{P}_o(A)$, a sufficient condition is that the maximum eigenvalue norm of $\bm{P}(A)$ is less than $\hf$.
If $\bm{P}(A)$ is of degree no more than $d$, then the circuit requires $d$ calls to $U$, $U^\dagger$ and their controlled gates in total.

\section{Applications}

\subsection{Bi-variate Quantum Signal Processing}\label{sec:mqsp}

Multi-variate quantum signal processing (M-QSP) generalizes uni-variate QSP to realize multi-variate polynomials $f(U_1,\cdots,U_n)$ using controlled operations on multiple signal operators.
While uni-variate QSP achieves polynomial transformations through alternating signal processing unitaries and signal operators, M-QSP must handle signal vectors $\bm{z} = (z_1,\ldots,z_m)$ where each component $z_k$ is encoded in a different signal operator, and the algorithm selects which operator to invoke at each step~\cite{rossi2022multivariable}.
The characterization of achievable polynomials in M-QSP has attracted considerable attention.
N\'emeth et al.\ established complete necessary and sufficient conditions for homogeneous and low-degree cases, but the general case remains open~\cite{nemeth2023variants}.
Laneve and Wolf provided the first general sufficient condition by extending to three-dimensional Hilbert space, requiring non-zero endpoint coefficients, though this condition is restrictive~\cite{laneve2025multivariate}.
Compared to the mature theory of uni-variate QSP, M-QSP lacks universal achievability criteria, with existing methods limited to specific assumptions such as variable commutativity, homogeneity, or low degree.

We propose an alternative approach based on the block encoding product rule.
Our method differs fundamentally from existing M-QSP frameworks: while previous approaches attempt to directly construct alternating product decompositions of multi-variate polynomials---requiring solutions to algebraic geometry problems such as multi-variate polynomial factorization and positive extension---our method transforms the problem into block encoding construction and linear combination of uni-variate polynomials, circumventing the difficulties of multi-variate decomposition.
Specifically, we express the bi-variate polynomial as $f(w,v) = \sum_j p_j(w)q_j(v)$, where $p_j(w)$ and $q_j(v)$ are uni-variate polynomials.
Using uni-variate $U(N)$-QSP, we construct block encodings $U_{p_j}$ and $V_{q_j}$ for $p_j(w)$ and $q_j(v)$ respectively, then obtain the block encoding of $p_j(w)q_j(v)$ through the block encoding product rule (\autoref{lem:prod}), and finally combine all terms through linear combination of unitaries~\cite{childs2012hamiltonian} to obtain the block encoding of $f(w,v)$.

We focus on bi-variate Laurent polynomials of the form
\begin{equation}
    f(\w,\v) = \sum_{j,k=-d}^{d} f_{jk} \w^j \v^k,
    \label{eq:bi-variate}
\end{equation}
where $\w$ and $\v$ are signal operators.
Throughout this section, we use $w,v\in\mathbb{C}$ with $|w|=|v|=1$ to denote complex scalars on the unit circle, and $\w,\v$ to denote signal operators that may be either scalars or commuting unitary matrices.
When $\w,\v$ are commuting unitaries, the polynomial $f(\w,\v)$ acts on their simultaneous eigenspaces, and our constructions apply with $\hat{w},\hat{v}$ representing the corresponding eigenvalues.
This is useful in scenarios like $\w=e^{i\kappa\q_1}$ and $\v=e^{i\kappa\q_2}$, where $\q_1$ and $\q_2$ are the position operators on two bosonic modes, to encode potential energy terms in Hamiltonian simulation.
We require all $\w$ to appear to the left of $\v$ and aim to realize the block encoding
\begin{equation}
    \begin{bmatrix}
        f(\w,\v) & * \\
        * & *
    \end{bmatrix}.
\end{equation}

\begin{figure*}
    \centering
    \includegraphics[width=0.6\textwidth]{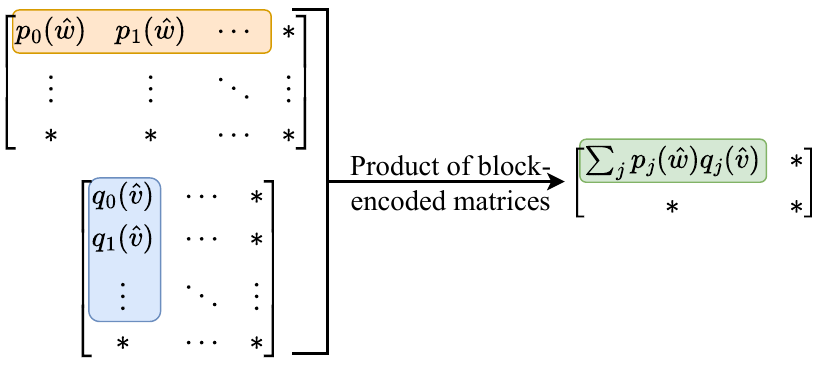}
    \caption{Block encoding of bi-variate polynomials via uni-variate $U(N)$-QSP. The product of block-encoded matrices is obtained by treating either the blue or orange box as a single block.}
    \label{fig:2_qsp}
\end{figure*}

Our protocol for constructing a block encoding of $f(\w,\v)$ is illustrated in \autoref{fig:2_qsp}. Two unitaries for variables $\w$ and $\v$ are constructed separately, and the block encoding of the target bi-variate polynomial is obtained via the following product rule.

\begin{lemma}[Product of block-encoded matrices~\cite{gilyen2019quantum}]\label{lem:prod}
    If $U$ and $V$ are block encodings of matrices $A$ and $B$ respectively, with ancilla qubits on different spaces $a$ and $b$, then $(I_b\otimes U)(I_a\otimes V)$ is a block encoding of the product $AB$.
\end{lemma}

To apply this lemma, we first express the polynomial as a linear combination of products of uni-variate polynomials:
\begin{equation}
    f(\w,\v) = \sum_{j} p_j(\w) q_j(\v),
    \label{eq:fwv}
\end{equation}
where $p_j(\w)$ and $q_j(\v)$ are uni-variate polynomials.
	The $U(N)$-QSP framework requires that the column vector $\{p_j(w)\}$ and row vector $\{q_j(v)\}$ have bounded norms: $\sum_j |p_j(w)|^2 \leq 1$ and $\sum_j |q_j(v)|^2 \leq 1$ for all unit complex numbers $|w|=|v|=1$.
	We call such bi-variate polynomials \emph{achievable} by the product of $U(N)$-QSP.

\begin{corollary}[Achievable polynomials]\label{cor:achievable}
    A bi-variate polynomial is achievable by product of $U(N)$-QSP if and only if it can be written as $f(\w,\v) = \sum_{j} p_j(\w) q_j(\v)$ such that $\sum_j |p_j(w)|^2 \leq 1$ and $\sum_j |q_j(v)|^2 \leq 1$ for all $|w|=|v|=1$.
\end{corollary}

While this condition provides a clear achievability criterion, finding a decomposition $f(\w,\v) = \sum_j p_j(\w) q_j(\v)$ satisfying the norm constraints requires numerical optimization.
A natural approach is to use low-rank decomposition of the coefficient matrix $\bm{F} = (f_{jk})_{j,k=-d}^{d}$.
However, direct decomposition may yield too many terms: for a $(2d+1) \times (2d+1)$ matrix, a full rank decomposition requires $O(d)$ terms, leading to correspondingly many $U(N)$-QSP sequences and excessive circuit complexity.

We employ principal component analysis (PCA) to find low-rank approximations.
Decomposing the coefficient matrix as
\begin{equation}\label{eq:pca}
    \bm{F} = \sum_{i=1}^{N} \lambda_i \bm{v}_i \bm{u}_i^\top,
\end{equation}
where $N = (2d+1)^2$, and retaining only the $r$ largest eigenvalues yields the rank-$r$ approximation
\begin{equation}\label{eq:rank-r-approx}
    \bm{F}_r = \sum_{i=1}^{r} \lambda_i \bm{v}_i \bm{u}_i^\top,
\end{equation}
corresponding to the polynomial approximation
\begin{equation}\label{eq:f-approx}
    f_r(\w,\v) = \sum_{i=1}^{r} \lambda_i p_i(\w) q_i(\v),
\end{equation}
where $p_i(\w) = \sum_{j=-d}^{d} (\bm{v}_i)_j \w^j$ and $q_i(\v) = \sum_{k=-d}^{d} (\bm{u}_i)_k \v^k$.
The reconstruction error is measured by
\begin{equation}\label{eq:recon-error}
    \varepsilon = \sum_{j,k} |f_{jk} - (f_r)_{jk}|.
\end{equation}

The effectiveness of PCA depends on the structure of the coefficient matrix.
In practical applications, target functions often possess smoothness, reflected in the decay of their Fourier coefficients.
Consider coefficient matrices with spatial decay
\begin{equation}\label{eq:decay}
    f_{jk} = r_{jk} \cdot w(j,k),
\end{equation}
where $r_{jk}$ is a random value and $w(j,k) = [(|j|+1)(|k|+1)]^{-s}$ with decay exponent $s > 0$.

\begin{figure*}
    \centering
    \includegraphics[width=.7\textwidth]{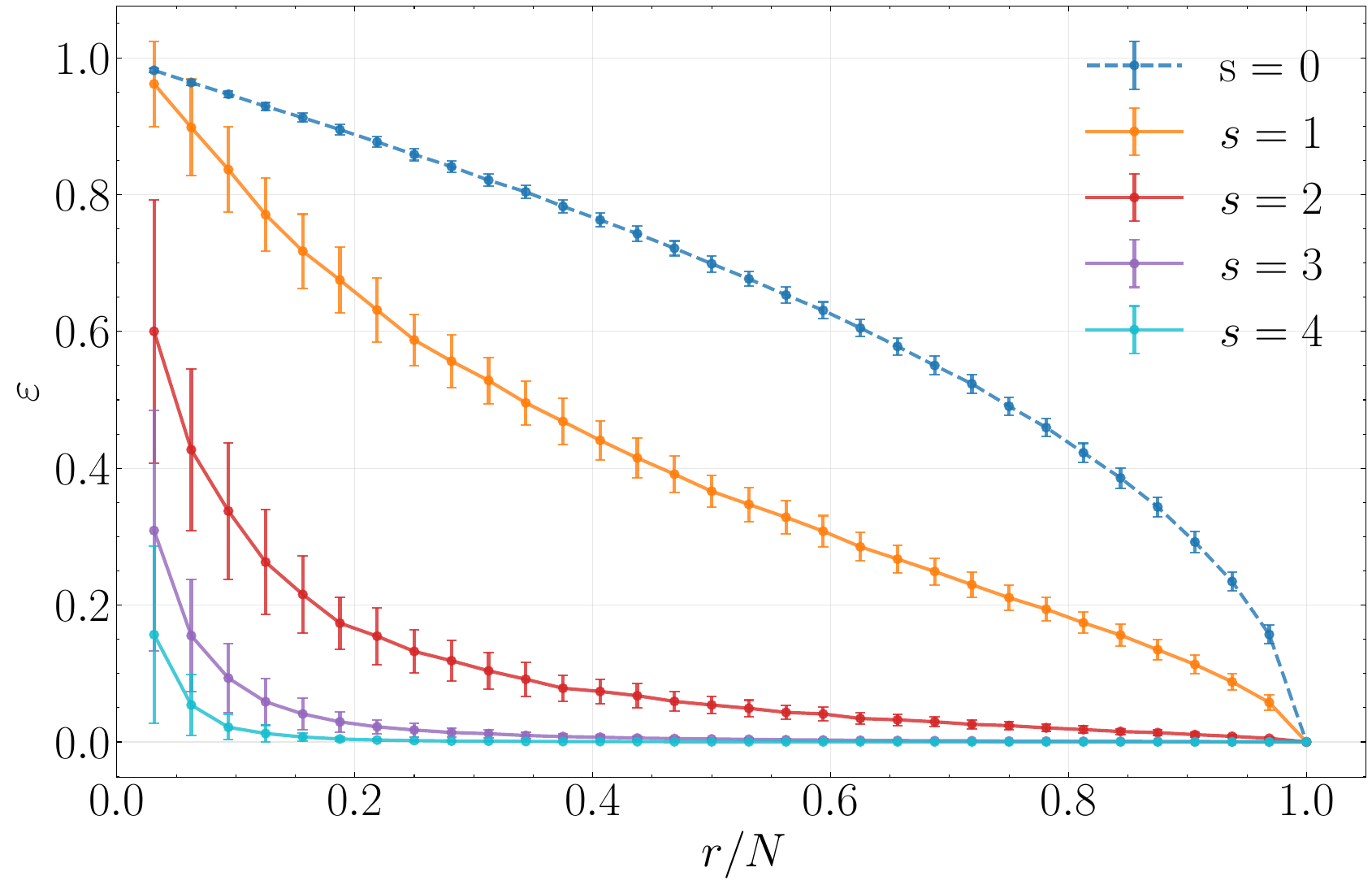}
    \caption{Comparison of PCA reconstruction error for different coefficient decay patterns. Polynomial decay significantly reduces the number of principal components needed to achieve the same error. Larger decay exponent $s$ yields better low-rank approximation. When $s=4$, only $r/N \approx 5\%$ is needed to achieve $\varepsilon < 0.1$, while random matrices require retaining over $90\%$ of principal components.}
    \label{fig:decay_comparison}
\end{figure*}

\autoref{fig:decay_comparison} compares the reconstruction error for different decay patterns at fixed matrix size $N=32$: no decay versus polynomial decay $w(j,k) = [(|j|+1)(|k|+1)]^{-s}$ with $s \in \{1, 2, 3, 4\}$.
For completely random coefficient matrices (no decay), the reconstruction error scales as approximately $\sqrt{1 - r/N}$ with the retained principal component ratio $r/N$, requiring about $99\%$ of principal components for small error ($\varepsilon < 0.1$).
In contrast, for polynomial decay with increasing exponent $s$, the required number of principal components decreases substantially: when $s=4$, only $r/N \approx 5\%$ is needed to achieve $\varepsilon < 0.1$.
This demonstrates that for functions with good smoothness properties, very few principal components suffice for high-accuracy reconstruction.

This result has significant implications for quantum algorithms.
In quantum simulation and optimization, target functions typically have physical meaning and thus possess smoothness and regularity, with Fourier coefficients exhibiting decay.
For such functions, even as the coefficient matrix dimension $d$ grows rapidly, PCA can keep the required rank $r$ small, significantly reducing the number of ancilla qubits and gate complexity in $U(N)$-QSP.

Compared to existing M-QSP methods, our approach offers several advantages.
First, the criterion is based on the complete theory of uni-variate QSP, avoiding the algebraic geometry difficulties of multi-variate positive extension without requiring verification of stability conditions or Fej\'{e}r-Riesz decomposition.
Second, the high-dimensional parameter space of the $U(N)$ framework makes numerical optimization more likely to find effective decompositions.
Third, for functions with regular structure, PCA provides a systematic approximation method whose effectiveness is guaranteed by the smoothness of the target function.

The method also has limitations.
While the condition is explicit, finding a satisfying decomposition still requires numerical optimization.
The method naturally handles commuting signal operators; for non-commuting unitary matrix inputs, the fixed variable ordering required by the product rule ($\w$ always before $\v$) restricts applicability to cases where this ordering is physically meaningful.
Additionally, the method requires $O(\log_2 r)$ ancilla qubits for linear combination, resulting in higher space complexity compared to existing methods that require only $1$ or $2$ ancilla qubits~\cite{rossi2022multivariable,laneve2025multivariate}.
Finally, the effectiveness of PCA strongly depends on the smoothness of the target function; for random or highly oscillatory functions, the rank reduction brings limited resource savings.

\subsection{Multi-interval Decision}\label{sec:decision}

\begin{figure*}
    \centering
    \includegraphics[width=.8\textwidth]{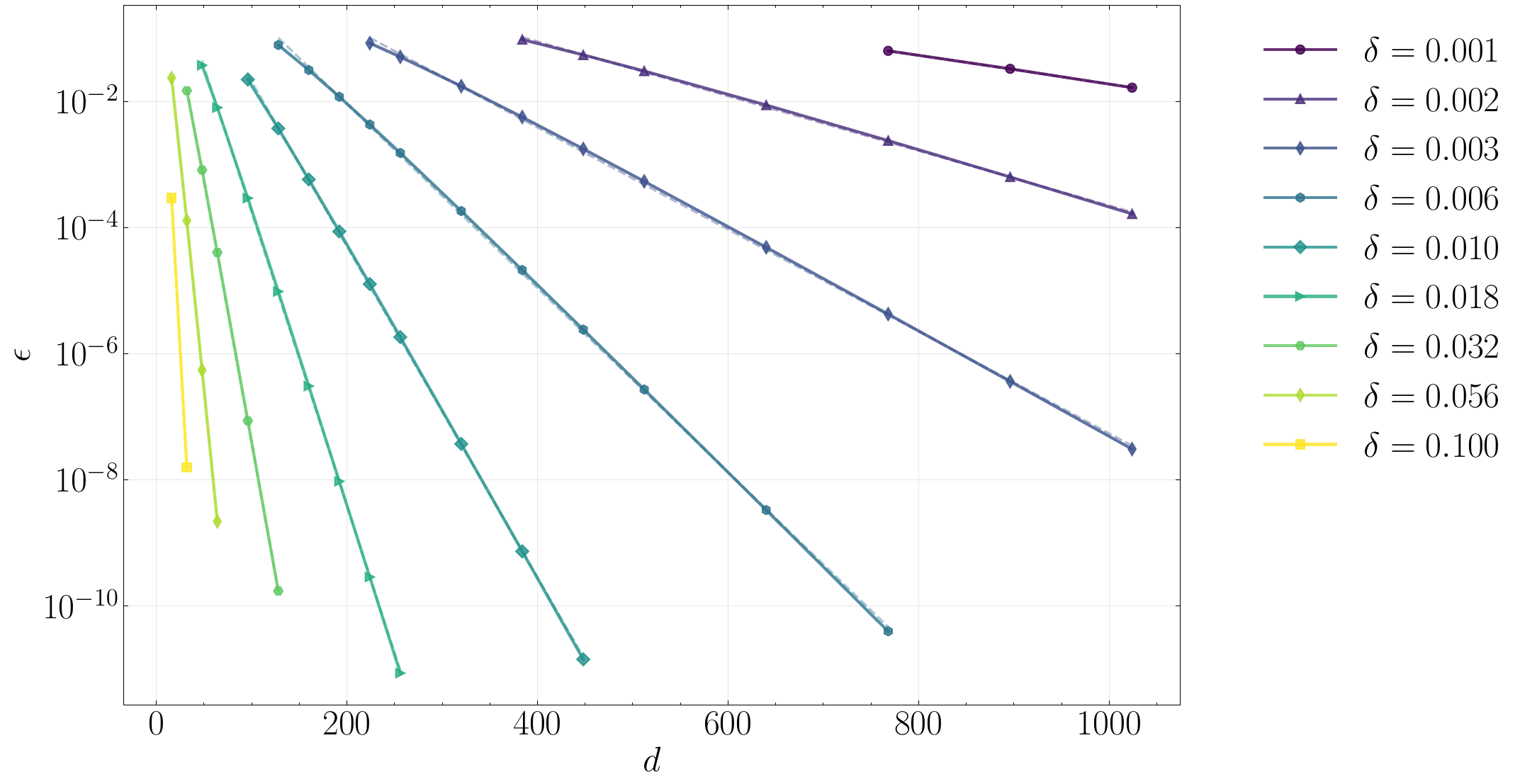}
    \caption{Empirical scaling of the degree $d$ required to achieve error $\epsilon$ for different gap widths $\delta$. The linear relationship between $d$ and $\log(\epsilon^{-1})$ for each $\delta$ confirms the scaling $d \sim C(\delta) \cdot \log(\epsilon^{-1})$. The dashed lines fit $d = C(\delta) \cdot \log(\epsilon^{-1})$. Numerical regression shows that $C(\delta) \approx 0.170 \cdot \delta^{-1}$.}
    \label{fig:eigenvalue_scaling}
\end{figure*}

The multi-interval decision problem asks to determine which of $N$ disjoint intervals a physical quantity of a quantum system belongs to.
This problem has broad applications in quantum computing, including determining the interval containing the ground state energy of a Hamiltonian, the amplitude of a quantum state, or the eigenvalue phase of a unitary operator.
Traditional $U(2)$-QSP uses a single ancilla qubit, which from an information-theoretic perspective extracts only one bit of information per measurement, requiring $\log_2 N$ measurement rounds to distinguish $N$ intervals.
In contrast, $U(N)$-QSP uses an $N$-dimensional ancilla system, extracting $\log_2 N$ bits of information in a single measurement and achieving a factor of $\log_2 N$ improvement in oracle complexity.

Let $\cc{I_0,\ldots,I_{N-1}}$ be $N$ disjoint intervals on $[-\hf,\hf]$ with pairwise distance at least $2\delta$.
Consider determining which interval the parameter $\varphi$ in the eigenvalue $e^{i 2\pi \varphi}$ of a unitary operator $U$ with respect to state $\ket{\psi}$ belongs to.
Suppose one uses a quantum circuit that queries the signal $d$ times and output a random variable $j\in\{0,1,\ldots,N-1\}$ with probability $Pr(j|\varphi)$ when the true parameter is $\varphi$, to decide which interval $\varphi$ belongs to, with target error $\epsilon$ such that,
\begin{equation}\label{eq:decision_err}
\begin{aligned}
	&
	Pr(j|\varphi) \ge 1 - \epsilon, \quad \forall \varphi \in I_j.
\end{aligned}
\end{equation}

By the Fej\'{e}r-Riesz theorem~\cite{fejer1916trigonometrische,riesz1916problem}, for any function $f(\varphi) = \sum_{j=-d}^{d} f_j e^{i 2\pi j \varphi}$ satisfying $f(\varphi) \ge 1$ for all $\varphi \in [-\pi, \pi]$, one can by its square root $g(\varphi) = \sum_{j=0}^{d} g_j e^{i 2\pi j \varphi}$ such that $|g(\varphi)|^2 = f(\varphi)$ for all $\varphi \in [-\pi, \pi]$.

For $N=2$, one can find a pair of $Pr(0|\varphi)$ and $Pr(1|\varphi)$ that approximate the indicator functions of the two intervals respectively, and the Fej\'{e}r-Riesz theorem guarantees the existence of their square roots $P_0(\varphi)$ and $P_1(\varphi)$.
Using generalized QSP (\autoref{thm:qsp_u}), one can realize a circuit with output state,
\begin{equation}
\begin{aligned}
	&
	\ket{0} P_0(U) \ket{\psi} + \ket{1} P_1(U) \ket{\psi},
\end{aligned}
\end{equation} 
so measuring the first qubit gives probabilities $Pr(j|\varphi) = \|P_j(U)\ket{\psi}\|^2 = Pr(j|\varphi)$ for $j=0,1$.

In the problem of multi-interval decision, one can use binary search by repeating the two-interval decision process $\log_2 N$ times with different circuits~\cite{martyn2021grand,sinanan2024single}.
Alternatively, we can construct a set of decision polynomials $P_j(\varphi)$ ($j=0,1,\ldots,N-1$) such that the measurement probabilities $Pr(j|\varphi) = |P_j(\varphi)|^2$ approximate the indicator functions of the $N$ intervals respectively.
To realize these polynomials, we need to use $U(N)$-QSP (\autoref{thm:unqsp_bk}) to implement the transformation
\begin{equation}
	\sum_{j=0}^{N-1} \ket{j} P_j(U) \ket{\psi}.
\end{equation}

This allows us to decide which interval $\varphi$ belongs to by measuring all the ancilla register in one go, saving a factor of $\log_2 N$ in oracle complexity compared to binary search.

In the rest of this section, we provide a numerical construction of decision polynomials $Pr(j|\varphi)$ that works for binary decision and general multi-interval decision, along with error bounds.

We define a family of Fourier polynomials optimized for decision problems.
Let
\begin{equation}\label{eq:qsp_poly_units}
    p_{d,\delta}(\varphi) = \abs{\sum_{j=0}^{d} c_j e^{i 2\pi j \varphi}}^2,
\end{equation}
and we minimize the window cost function,
\begin{equation}\label{eq:def_window_cost}
\begin{aligned}
	W_{\delta}
	= &
	\int_{-1/2}^{1/2} w_\delta(\varphi) p_{d,\delta}(\varphi) d\varphi
	\\ = &
	\sum_{j,k=0}^{d} c_j c_k^* \int_{-1/2}^{1/2} w_\delta(\varphi) e^{i 2\pi (j-k) \varphi} d\varphi
	\\ = &
	\sum_{j,k=0}^{d} c_j c_k^* M_{jk},
\end{aligned}
\end{equation}
under constraints,
\begin{equation}\label{eq:cstr_unit_poly}
\begin{aligned}
	&
	\int_{-1/2}^{1/2} p_{d,\delta}(\varphi) d\varphi
	=
	\sum_{j=0}^{d} \abs{c_j}^2
	= 1,
\end{aligned}
\end{equation}
where $w_{\delta}(\varphi) = 0$ for $\varphi \in [-\delta,\delta]$ and $1$ elsewhere on $[-1/2,1/2]$, and
\begin{equation}\label{eq:mjk}
\begin{aligned}
	M_{jk}
	= &
	\int_{-1/2}^{1/2} w_\delta(\varphi) e^{i 2\pi (j-k) \varphi} d\varphi
	\\ = &
	\begin{cases}
		1 - 2\delta, & j=k, \\
		-\frac{\sin(2\pi (j-k) \delta)}{\pi (j-k)}, & j \ne k.
	\end{cases}
\end{aligned}
\end{equation}

The minimal $W_{\delta}$ can be obtained by choosing $\cc{c_j}$ as the eigenvector corresponding to the minimal eigenvalue of the $(N+1)\times(N+1)$ matrix $M = (M_{jk})$.
To achieve window cost $\epsilon$ for fixed $\delta$, one needs a large value of $d$ such that a more narrow peak can be formed within $[-\delta,\delta]$.
We numerically compute $\epsilon$ for various $d$ and $\delta$, and find the following empirical scaling relation (\autoref{fig:eigenvalue_scaling}).

\begin{claim}[Window cost scaling]\label{clm:delta_scaling}
	For the numerically optimized Fourier polynomial $p_{d,\delta}$ defined in \autoref{eq:qsp_poly_units}, in the regime $\delta \in [10^{-3}, 10^{-1}]$, the polynomial degree required to achieve error $\epsilon$ for optimal window cost satisfies
	\begin{equation}\label{eq:delta_scaling}
		d \gtrsim 0.170 \cdot \delta^{-1} \cdot \log(\epsilon^{-1}).
	\end{equation}
\end{claim}

The following theorem provides an explicit construction and error bound for general multi-interval decision.

\begin{theorem}[Degree in multi-interval decision]\label{thm:mdp_poly}
    Let intervals $I_j=[a_j,b_j]$ ($j=0,1,\ldots,N-1$) be disjoint with pairwise distance $2\delta$.
    For each interval $I_j$, define the decision polynomial
    \begin{equation}\label{eq:mdp_poly}
        Pr(j|\varphi) = \int_{\tilde{I}_j} p_{d,\delta}(\varphi - \varphi') d\varphi',
    \end{equation}
    where $\tilde{I}_j := [a_j-\delta, b_j+\delta]$ is the extended interval and $p_{d,\delta}$ is defined in \autoref{eq:qsp_poly_units}.
	Then, in the regime $\delta \in [10^{-3}, 10^{-1}]$, one can achieve
    \begin{equation}\label{eq:mdp_err_bound}
		Pr(j|\varphi) \ge 1 - \epsilon, \quad \forall j \text{ and } \varphi \in I_j.
    \end{equation}
	with
    \begin{equation}\label{eq:mdp_degree}
        d \gtrsim 0.170 \cdot \delta^{-1} \cdot \log(\epsilon^{-1})
    \end{equation}
	number of queries to the signal.
\end{theorem}

\pf{thm:mdp_poly}
For any $\varphi\in I_j$ we have,
\begin{equation}
\begin{aligned}
	&
	1 - Pr(j|\varphi)
	\\ = &
	\int_{[-\pi,\pi] - \tilde{I}_j} p_{d,\delta}(\varphi - \varphi') d\varphi'
	\\ \le &
	\int_{[-\pi,\pi] - [-\delta,\delta]} p_{d,\delta}(\varphi - \varphi') d\varphi'
	\\ = &
	\epsilon,
\end{aligned}
\end{equation}
therefore, one can refer to \autoref{eq:delta_scaling} to determine the required degree $d$ to achieve error $\epsilon$.
\qed

\begin{figure*}
    \centering
    \subfloat[4-decision polynomials for $U(N)$-QSP using numerically optimized Fourier polynomials ($d=16$, $\delta=0.1$). The four polynomials $P_0, P_1, P_2, P_3$ each approximate $1$ in their corresponding intervals, demonstrating one-shot multi-interval decision capability of $U(N)$-QSP.\label{fig:qsp_quatro_decision}]{%
        \includegraphics[width=.48\textwidth]{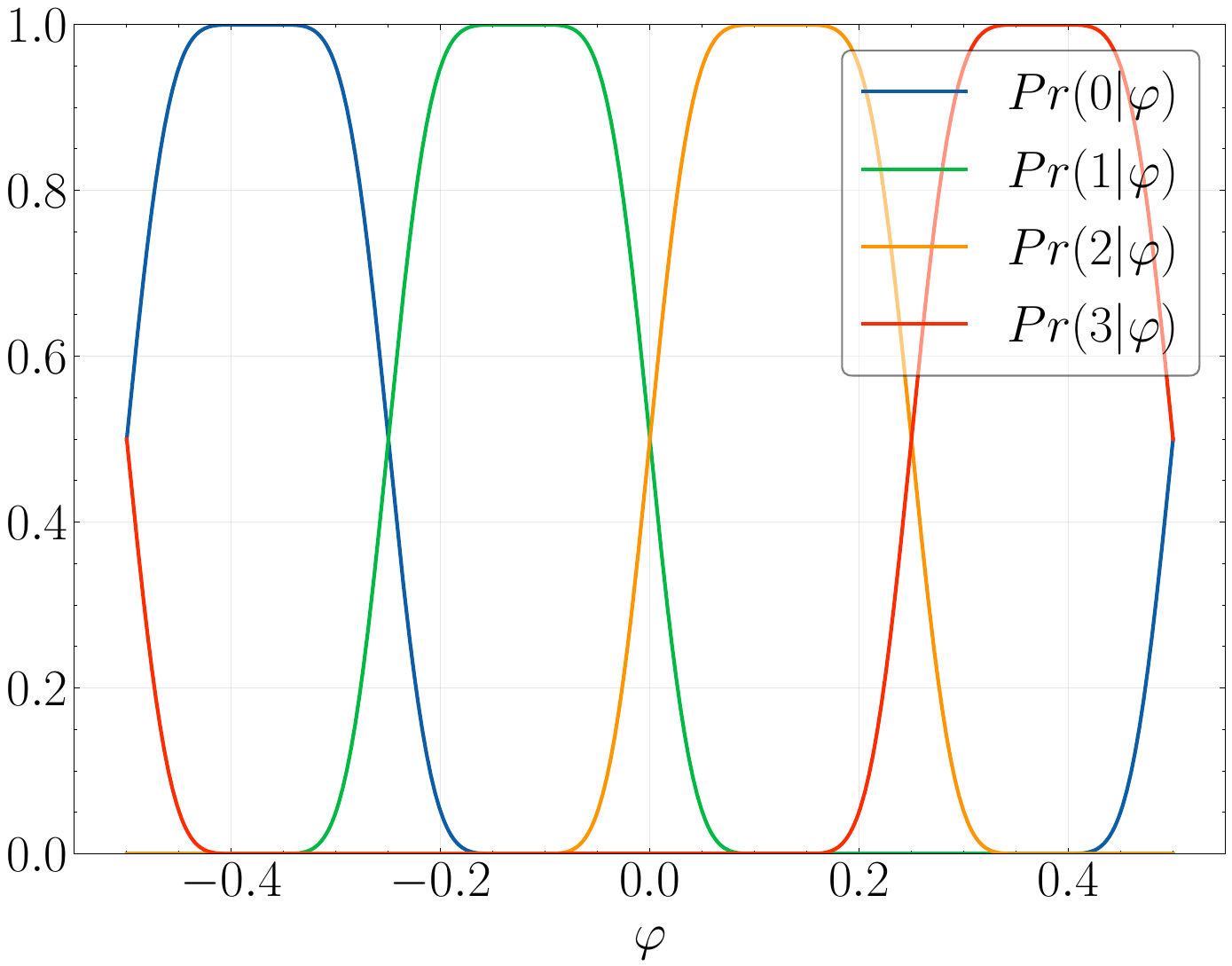}%
    }
	\quad
    \subfloat[Comparison of $U(4)$-QSP direct method versus $U(2)$-QSP two-stage method for 4-decision. Solid lines show $U(4)$-QSP, dashed lines show $U(2)$-QSP. The $U(4)$-QSP method achieves the same error with roughly half the number of queries.\label{fig:qsp_quatro_comparison}]{%
        \includegraphics[width=.46\textwidth]{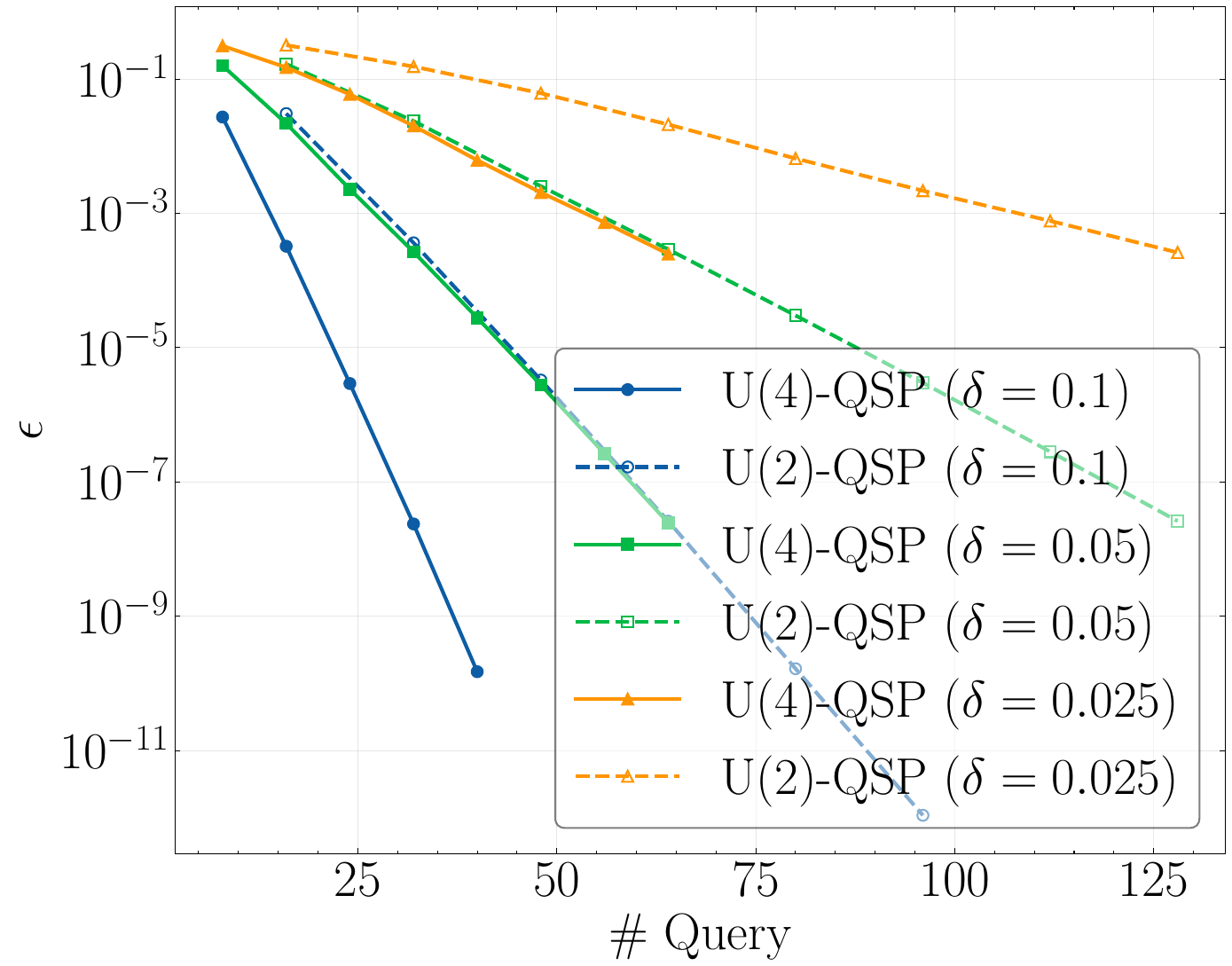}%
    }
    \caption{4-decision QSP polynomials and comparison.}
    \label{fig:qsp_quatro}
\end{figure*}

\textit{Example: 4-interval decision.---}
As a concrete example, we consider the 4-decision problem where we determine whether $\varphi$ belongs to $I_0=[-\pi+\delta,-\frac{\pi}{2}-\delta]$, $I_1=[-\frac{\pi}{2}+\delta,-\delta]$, $I_2=[\delta,\frac{\pi}{2}-\delta]$, or $I_3=[\frac{\pi}{2}+\delta,\pi-\delta]$.
\autoref{fig:qsp_quatro_decision} shows the 4 decision polynomials for $d=16$ and $\delta=0.1$ by \autoref{eq:mdp_poly}, demonstrating that each polynomial approximates $1$ in its corresponding interval and $0$ elsewhere.
One can implement these polynomials using $U(4)$-QSP with $d=16$ queries to the signal in one go.

An alternative approach using traditional $U(2)$-QSP would require two-stage binary decision: the first stage performs binary decision at $\varphi=0$ to distinguish left and right half-planes, and the second stage performs binary decisions at $\varphi=-\frac{\pi}{2}$ and $\varphi=\frac{\pi}{2}$ respectively.
This $U(2)$-QSP based method extracts only one bit of information per stage, requiring $\log_2 N$ rounds.
\autoref{fig:qsp_quatro_comparison} compares the error performance of both approaches, clearly showing that $U(4)$-QSP achieves the same error with roughly half the number of signal calls.

\subsection{Quantum Amplitude Estimation}\label{sec:qae}

Quantum amplitude estimation (QAE) addresses the following problem: given a state preparation operator $U$ that prepares $\ket{\psi}=U\ket{\psi_0}$ from an initial state $\ket{\psi_0}$, and a projection operator $\Pi$, estimate $x=\ev{\Pi}{\psi}$ using $N$ queries to $U$ and $U^{-1}$.

This is useful in the task of estimating the expectation value of an observable $A$ with respect to a state $\ket{\psi}$, where we assume $A$ has all eigenvalues in $[-1,1]$ and is block-encoded by $U$.
The expectation value, $\langle\psi|A|\psi\rangle$, of a block-encoded observable $A$ can be estimated via quantum amplitude estimation if we consider the state $(|0\rangle|\bm{0}\rangle|\psi\rangle + |0\rangle U|\bm{0}\rangle|\psi\rangle)/\sqrt{2}$ and the projector $|+\rangle\langle +| \otimes I$, where the identity acts on the block-encoding ancilla and system qubits.

Let $Pr(k|x)$ denote the probability of obtaining the $k$-th measurement outcome when the amplitude is $x$.
We assume the circuit has a fixed structure that does not adaptively change based on intermediate results, and each output probability $Pr(k|x)$ depends only on $x$.
The total number of calls to $U$ and $U^{-1}$ is the \emph{degree} of the QAE circuit, denoted $d$.

\begin{lemma}[QAE probabilities are polynomials]\label{lem:poly}
    Each output probability of a degree-$d$ QAE circuit is a polynomial in $x$ of degree at most $d$.
\end{lemma}

\noindent\textit{Proof.}
Consider a simple QAE setup with state preparation
\begin{equation}
\begin{aligned}
    W(\theta) = \begin{bmatrix} \cos(\theta/2) & -\sin(\theta/2) \\ \sin(\theta/2) & \cos(\theta/2) \end{bmatrix},
\end{aligned}
\end{equation}
initial state $\ket{0}$, and projector $\dyad{0}$, where the target amplitude is $x=\cos^2(\theta/2)$.
By induction on $d$, after $d$ calls to $W(\theta)$ and its inverse, the quantum state becomes a polynomial vector in $\cos(\theta/2)$ and $\sin(\theta/2)$ of degree at most $d$ with definite parity.
Any projection measurement probability must have the form $P_1(x) + \sin\theta\, P_2(x)$, where $P_1$ and $P_2$ are polynomials of degrees at most $d$ and $d-1$ respectively.
However, since $W(\theta)$ and $W(-\theta)$ share the same amplitude $x$ but opposite sign of $\sin\theta$, and must yield identical probabilities, we conclude $P_2 = 0$, establishing that the probability is indeed a polynomial in $x$.
\qed

As a concrete example, if we apply the amplitude amplification operator~\cite{brassard2000quantum}
\begin{equation}\label{eq:amp-amplification}
    U(I-2\dyad{\psi_0})U^{-1}(I-2\Pi)
\end{equation}
$k$ times to $U\ket{\psi_0}$ and measure on $\{\Pi,I-\Pi\}$, the output probability for $\Pi$ is
\begin{equation}\label{eq:amp-prob}
    \sin^2\left(\frac{2k+1}{2}\theta\right) = \frac{1-T_{2k+1}(2x-1)}{2},
\end{equation}
a polynomial of degree $2k+1$ in $x$, where $T_k$ is the Chebyshev polynomial of the first kind.
This matches the QAE circuit degree since each amplification adds degree 2, plus 1 for initial state preparation, giving total degree $d=2k+1$.

QAE is structurally similar to multi-interval decision.
Given target precision $\epsilon$, partition $[0,1]$ into $k=\mathcal{O}(\epsilon^{-1})$ intervals $\{I_j = [j\epsilon, (j+1)\epsilon]\}$.
The goal is to determine which interval contains the unknown amplitude $x$, thereby estimating $x$ with precision $\mathcal{O}(\epsilon)$.
This shares the same structure as multi-interval decision discussed in \autoref{sec:decision}, but with each sampling point directly corresponding to a measurement outcome rather than being integrated over an interval.

\begin{figure*}
    \centering
    \subfloat[Measurement outcome probabilities $Pr(k|x)$ for QAE with $N=16$.]{%
        \includegraphics[width=.48\textwidth]{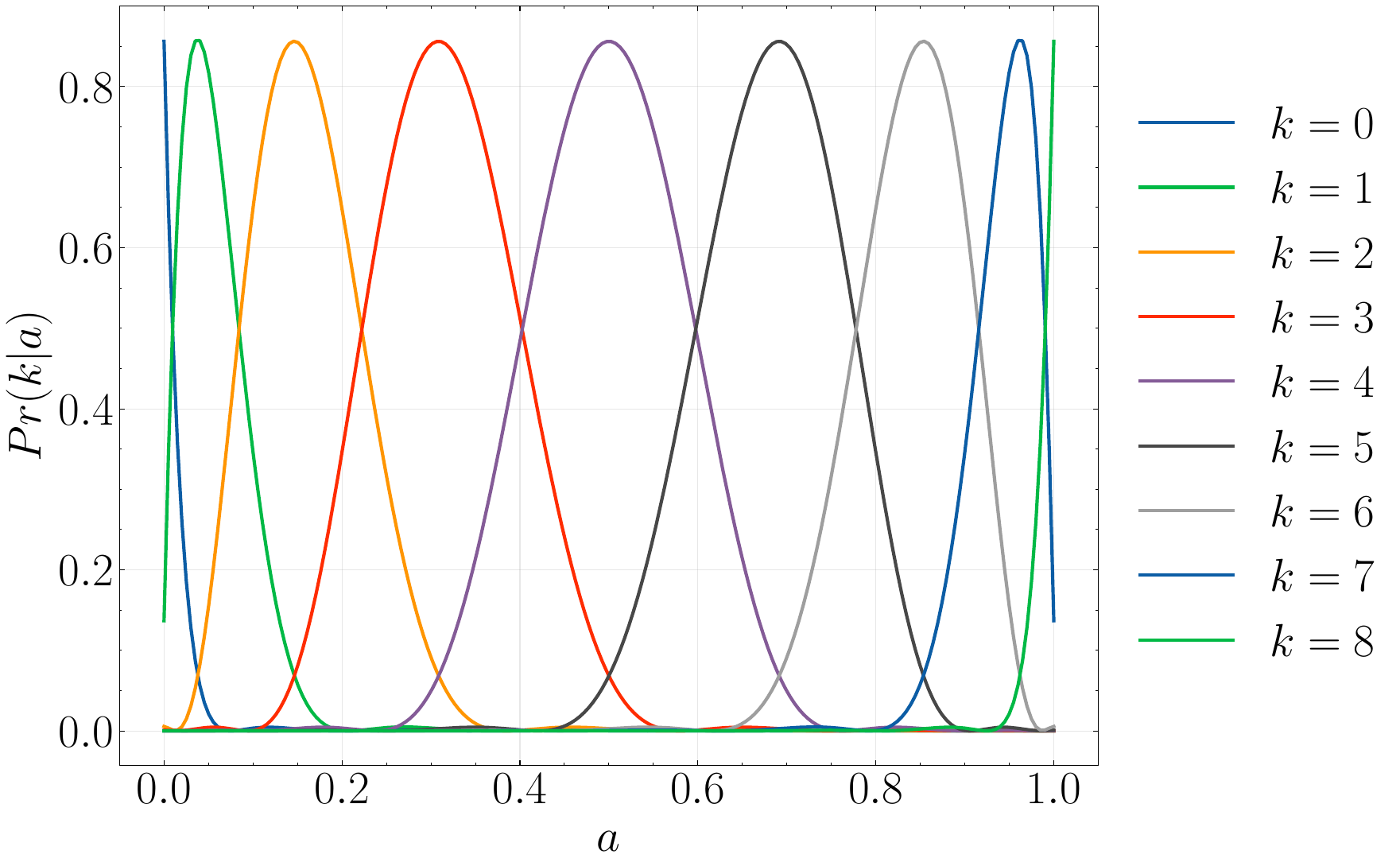}%
		\label{fig:qae_probs}
    }
    \quad
    \subfloat[RMSE versus circuit degree $N$ for QAE.]{%
        \includegraphics[width=.46\textwidth]{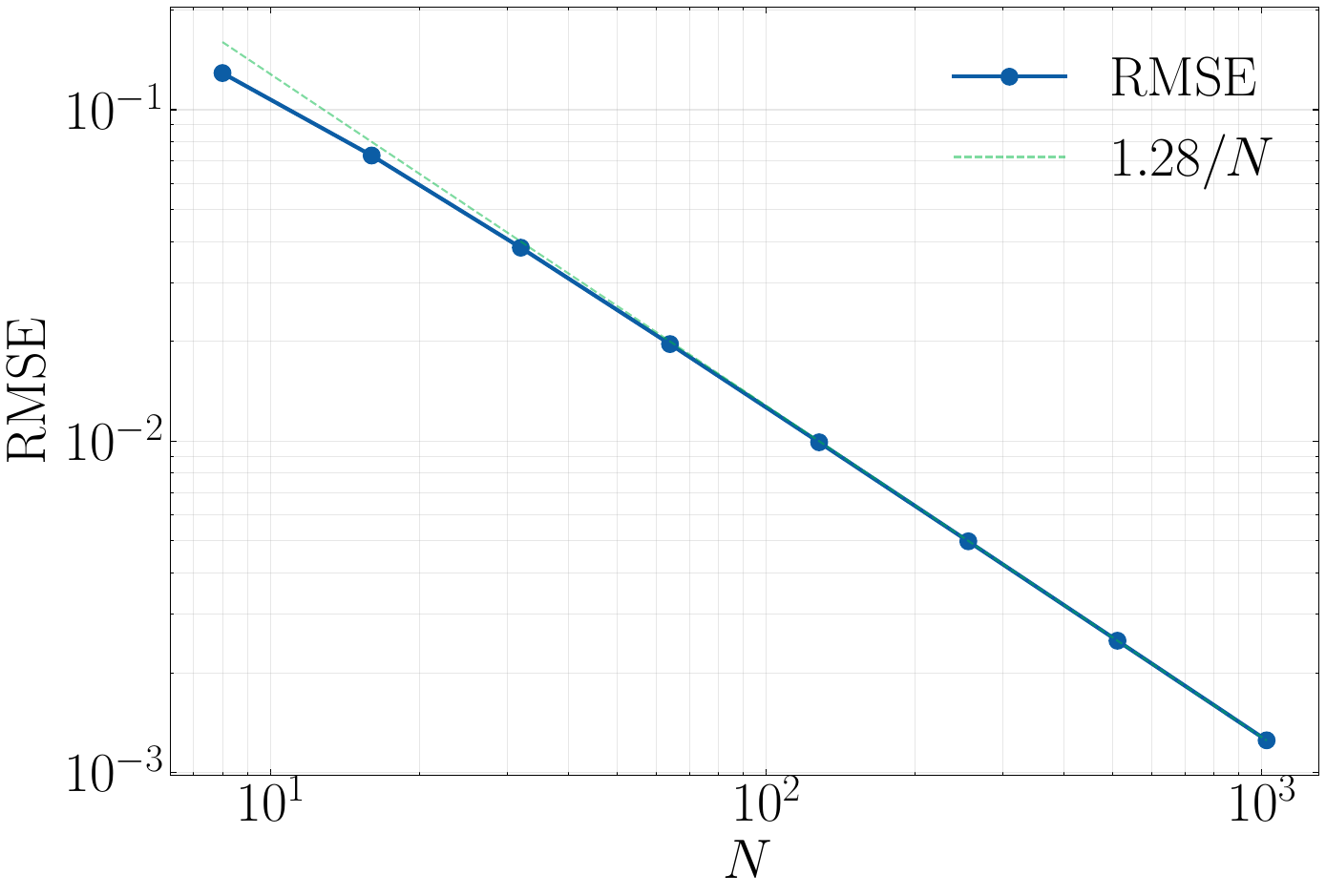}%
		\label{fig:qae_rmse}
    }
    \caption{QAE measurement probabilities and RMSE scaling.}
    \label{fig:qae}
\end{figure*}

\begin{theorem}[QAE via $U(N)$-QSVT]\label{thm:qae-eqv}
    For any set of probability functions $\{Pr(k|x)\}$ that are polynomials in $x$ of degree at most $d$, non-negative on $[0,1]$, and satisfy $\sum_k Pr(k|x) \equiv 1$, there exists a $U(N)$-QSVT circuit that produces measurement outcome $k$ with probability $Pr(k|x)$.
\end{theorem}

\noindent\textit{Proof.}
Using Lemma 6 of \cite{gilyen2019quantum}, we decompose each probability polynomial as
\begin{equation}\label{eq:poly-decomp}
\begin{aligned}
	&		
	Pr(k|\cos^2(\theta/2))
	\\ = &
	A_k(\cos(\theta/2))^2 + \sin^2(\theta/2) B_k(\cos(\theta/2))^2
\end{aligned}
\end{equation}
for suitable polynomials $A_k$ and $B_k$ of degrees $d$ and $d-1$ respectively.
Write the state preparation as
\begin{equation}\label{eq:decomp-tH}
    U\ket{\psi_0} = \cos(\theta/2)\ket{\tpsi_0} + \sin(\theta/2)\ket{\tpsi_1},
\end{equation}
where $\ket{\psi_0}$ is on the projected subspace of $\Pi$, and $\ket{\psi_1}$ is orthogonal to $\ket{\psi_0}$.
Define $\ket{\psi_1} = U^{-1}[-\sin(\theta/2)\ket{\tpsi_0} + \cos(\theta/2)\ket{\tpsi_1}]$ orthogonal to $\ket{\psi_0}$.
In the basis $(\ket{\psi_0},\ket{\psi_1})\to(\ket{\tpsi_0},\ket{\tpsi_1})$, $U$ has matrix representation
\begin{equation}\label{eq:u-basis-repr}
    U = \begin{bmatrix}
        \cos(\theta/2) & -\sin(\theta/2) \\
        \sin(\theta/2) & \cos(\theta/2)
    \end{bmatrix}.
\end{equation}
The target state for $U(N)$-QSVT is
\begin{equation}\label{eq:target-state}
\begin{aligned}
	\sum_{k=0}^{N-1} \ket{k} & \biggl[ A_k(\cos(\theta/2))\ket{\tpsi_0}
	\\
	+ & \sin(\theta/2) B_k(\cos(\theta/2))\ket{\tpsi_1}\biggr],
\end{aligned}
\end{equation}
if $d$ is odd (or with $\ket{\psi_0},\ket{\psi_1}$ replacing $\ket{\tpsi_0},\ket{\tpsi_1}$ if $d$ is even), which satisfies the conditions of \autoref{thm:unqsvt_bk}.
\qed

To minimize the standard deviation of the estimated amplitude, we use the decision polynomial,
\begin{equation}\label{eq:amp_kth}
    Pr(k|x) = \begin{cases}
        |p_{\sin}(\theta_x-\theta_k)|^2 + |p_{\sin}(\theta_x+\theta_k)|^2, \\
        \qquad\qquad\qquad k=1,\ldots,\frac{N}{2}-1, \\
        |p_{\sin}(\theta_x-\theta_k)|^2, k=0,\frac{N}{2},
    \end{cases}
\end{equation}
where $\theta_x = \arccos(1-2x)$, $\theta_k = k\pi/N$ for $k=0,1,\ldots,N/2$, and
\begin{equation}\label{eq:sine_poly}
\begin{aligned}
	&
	p_{\sin}(\varphi) = \sqrt{\frac{2}{N(N+1)}} \sum_{k=0}^{N-1} \sin\left(\frac{(k+1)\pi}{N+1}\right) e^{ik\varphi},
\end{aligned}
\end{equation}
which is known to minimize the variance in phase estimation~\cite{van2007a,van2007b,gorecki2020pi}.
The range of $k$ excludes values greater than $N/2$ because they correspond to amplitudes already covered by $k \le N/2$ due to the symmetry $\theta_x \in [0,\pi]$.
Using Bayesian estimation, the amplitude estimate for outcome $k$ is
\begin{equation}\label{eq:amp_est}
    x_k = \frac{\int_0^1 x Pr(k|x) dx}{\int_0^1 Pr(k|x) dx}.
\end{equation}

\autoref{fig:qae_probs} shows the probability curves $Pr(k|x)$ for $N=8$.
Different outcomes $k$ have high probability in different amplitude regions, enabling inference of the amplitude from the measurement result.

To analyze performance, we compute the root mean square error (RMSE)
\begin{equation}\label{eq:qae_rmse}
    \mathrm{RMSE} = \sqrt{\int_0^1 \sum_{k=0}^{N/2} (x_k - x)^2 Pr(k|x) dx},
\end{equation}
where $x_k$ is given by \autoref{eq:amp_est}.
\autoref{fig:qae_rmse} shows RMSE versus circuit degree $N$.
The numerical results indicate RMSE decreases as $\mathcal{O}(N^{-1})$, matching the reference line $1.28 N\inv$.
This demonstrates that the $U(N)$-QSVT based QAE achieves the Heisenberg limit, where estimation precision is inversely proportional to the circuit degree.
Similarly, $U(2)$-based QAE requires $\mathcal{O}(\log_2 N)$ rounds of binary search to achieve the same precision~\cite{sinanan2024single,martyn2021grand}.

\section{Conclusion and Outlook}\label{sec:conclusion}

We have established a comprehensive theoretical framework for quantum signal processing and quantum singular value transformation on $U(N)$, generalizing the standard $U(2)$ framework by using $N$-dimensional ancilla systems with parameterized unitary operations.
Our main theoretical contribution is a constructive characterization of achievable polynomial matrices: any polynomial matrix $\bm{P}(z)$ satisfying appropriate norm constraints on the unit circle can be block-encoded using controlled-$U$ operations, with explicit recursive algorithms provided for circuit parameter determination.

Three applications demonstrate the practical advantages of this framework.
First, we present a method for bi-variate quantum signal processing that leverages the product rule for block encodings and principal component analysis, avoiding the algebraic difficulties of existing multi-variate QSP approaches.
Second, we show that $U(N)$-QSP achieves multi-interval decision with oracle complexity $\mathcal{O}(d)$ for $N$ intervals, achieving a factor of $\log_2 N$ improvement over the $\mathcal{O}(d\log_2 N)$ complexity required by iterative $U(2)$-QSP methods.
Third, we establish that quantum amplitude estimation via $U(N)$-QSVT achieves the Heisenberg limit scaling $\mathcal{O}(N^{-1})$ in a single non-adaptive measurement, eliminating the need for the $\mathcal{O}(\log_2 N)$ adaptive rounds required in conventional approaches.

Several directions for future research are promising.
First, developing efficient classical algorithms for computing the unitary parameters $R_0, \ldots, R_d$ beyond the inductive construction, potentially using techniques from numerical optimization or tensor networks, would improve practical applicability.
Second, investigating whether the expanded parameter space of $U(N)$-QSP can enhance the expressivity of variational quantum algorithms compared to $U(2)$-based approaches could reveal new applications in quantum machine learning.
Third, extending PCA-based decomposition methods for bi-variate QSP to higher-order multivariate polynomials and exploring connections to tensor decomposition would address the broader challenge of multi-variate quantum signal processing.
Finally, practical implementation of $U(N)$-QSP on real quantum hardware, studying the effects of decoherence and gate errors on circuit fidelity, and extending these techniques to non-qubit architectures such as bosonic systems and hybrid quantum processors~\cite{sinanan2024single,liu2024hybrid} remain important open questions.

\section*{Acknowledgement}

We thank Bojko Bakalov for fruitful discussions.
X.L. and H.L. acknowledge support from the National Natural Science Foundation of China under Grant Nos.~62272406, 61932018.

\section*{Code Availability}

The source code for numerical experiments and implementations presented in this paper is available at \url{https://github.com/helloluxi/unqsp-code}.

\bibliography{ref.bib}
\bibliographystyle{unsrturl}

\onecolumn
\appendix

\end{document}